\documentclass[a4paper]{article}

\usepackage{graphicx}
\usepackage[space]{grffile}
\usepackage{latexsym}
\usepackage{textcomp}
\usepackage{longtable}
\usepackage{tabulary}
\usepackage{booktabs,array,multirow}
\usepackage{amsfonts,amsmath,amssymb}
\usepackage{url}
\usepackage[colorlinks,bookmarksnumbered,citecolor=red,urlcolor=red]{hyperref}
\usepackage{etoolbox}

\usepackage{amsmath,amssymb}
\usepackage{algorithm}
\usepackage{algorithmic}
\usepackage{multirow}
\usepackage{tikz}
\usepackage{soul}
\usepackage{bm}
\usepackage{subfig}
\usetikzlibrary{patterns}
\newcommand*{\ditto}{---\texttt{"}---}
\newcommand\mapsfrom{\mathrel{\reflectbox{\ensuremath{\mapsto}}}}
\renewcommand{\vec}[1]{\ensuremath{\boldsymbol{#1}}}
\newcommand{\figdir}{./}
\newcommand{\hatchingNW}{\begin{tikzpicture}\draw[pattern=north west lines, pattern color=black] (0,0) rectangle (0.2,0.2);\end{tikzpicture}}
\newcommand{\hatchingNE}{\begin{tikzpicture}\draw[pattern=north east lines, pattern color=black] (0,0) rectangle (0.2,0.2);\end{tikzpicture}}
\newcommand{\Tmixed}{\ensuremath{T_{\text{solve}}^{(\text{m})}}}
\newcommand{\Tpress}{\ensuremath{T_{\text{solve}}^{(\text{p})}}}
\newcommand{\tmixed}{\ensuremath{t_{\text{solve}}^{(\text{m})}}}
\newcommand{\tpress}{\ensuremath{t_{\text{solve}}^{(\text{p})}}}

\usepackage[utf8]{inputenc}
\usepackage[english]{babel}
\usepackage[cm]{fullpage}
\usepackage{amsthm}
\usepackage{multicol}
\usepackage{graphicx}
\usepackage[affil-it]{authblk}

\usepackage{siunitx}

\begin{document}
\author[a,b]{Christopher~Maynard}
\author[a]{Thomas~Melvin}
\author[c,*]{Eike~Hermann~M\"{u}ller}
\affil[a]{Met Office, Fitzroy Rd, Exeter EX1 3PB, U.K.}
\affil[b]{Department of Computer Science, University of Reading, Reading RG6 6AY, U.K.}
\affil[c]{Department of Mathematical Sciences, University of Bath, Bath BA2 7AY, Bath, U.K.}
\affil[*]{Email: \texttt{e.mueller@bath.ac.uk}}

\title{Multigrid preconditioners for the mixed finite element dynamical core of the LFRic atmospheric model}
\makeatletter
\def\blfootnote{\gdef\@thefnmark{}\@footnotetext}
\makeatother
\blfootnote{\textcopyright \hspace{0.5mm} Crown copyright 2020. Reproduced with the permission of the Controller of HMSO.}


\maketitle
\selectlanguage{english}
\begin{abstract}\noindent
Due to the wide separation of time scales in geophysical fluid dynamics, semi-implicit time integrators are commonly used in operational atmospheric forecast models. They guarantee the stable treatment of fast (acoustic and gravity) waves, while not suffering from severe restrictions on the timestep size. To propagate the state of the atmosphere forward in time, a non-linear equation for the prognostic variables has to be solved at every timestep. Since the nonlinearity is typically weak, this is done with a small number of Newton- or Picard- iterations, which in turn require the efficient solution of a large system on linear equations with $\mathcal{O}(10^6-10^9)$ unknowns. This linear solve is often the computationally most costly part of the model. In this paper an efficient linear solver for the LFRic next-generation model, currently developed by the Met Office, is described. The model uses an advanced mimetic finite element discretisation which makes the construction of efficient solvers challenging compared to models using standard finite-difference and finite-volume methods. The linear solver hinges on a bespoke multigrid preconditioner of the Schur-complement system for the pressure correction. By comparing to Krylov-subspace methods, the superior performance and robustness of the multigrid algorithm is demonstrated for standard test cases and realistic model setups. In production mode, the model will have to run in parallel on 100,000s of processing elements. As confirmed by numerical experiments, one particular advantage of the multigrid solver is its excellent parallel scalability due to avoiding expensive global reduction operations.
\\[1ex]
\end{abstract}%


\section{Introduction}
Operational models for numerical climate- and weather-prediction have to solve the equations of fluid dynamics in a very short time. State-of-the art implementations rely on accurate spatial discretisations and efficient timestepping algorithms. To make efficient use of modern supercomputers they have to exploit many levels of parallelism (such as SIMD/SIMT, threading on a single node and message passing on distributed memory systems) and scale to 100,000s of processing elements. Semi-implicit time integrators are commonly employed since they allow the stable treatment of fast acoustic waves, which carry very little energy but have to be included in a fully compressible formulation. The implicit treatment of the acoustic modes allows the model to run with a relatively long timestep. The size of the timestep is only restricted by advection which horizontally is generally around one order of magnitude slower than acoustic oscillations and vertically two orders of magnitude slower. The main computational cost of semi-implicit models is the repeated solution of a very large sparse system of linear equations. While standard iterative solution algorithms exist, the linear system is ill-conditioned, which leads to the very slow convergence of Krylov-subspace methods. This is a particularly serious problem for massively parallel implementations due to their very large number of global reduction operations (arising from vector dot-products and norms in each Krylov-iteration). Preconditioners, which solve an approximate version of the linear system, overcome this issue and dramatically reduce the number of Krylov iterations and global communications. The construction of an efficient preconditioner is non-trivial and requires careful exploitation of the specific properties of the system which is to be solved. For global atmospheric models, two key features which have to be taken into account are (1) the high aspect ratio arising from the shallow domain and (2) the finite speed of sound in compressible formulations, which limits the effective distance over which different points in the domain are correlated during one timestep. Typically the preconditioner is based on a Schur-complement approach. This reduces the problem to an elliptic equation for the pressure correction, which can then be solved with standard methods.
\paragraph{Multigrid.}
Hierarchical methods such as multigrid algorithms (\cite{Trottenberg2000}) are often employed for the solution of elliptic systems since they have a computational complexity which grows linearly with the number of unknowns. \cite{Mueller2014} contains a recent review of linear solvers in atmospheric modelling (see also \cite{Steppeler2003}). There, the performance of a multigrid solver based on the tensor-product algorithm described in \cite{Borm2001} was applied to a simplified model system which is representative for the linear system for the pressure correction. The key idea is to use a vertical line-relaxation smoother together with semi-coarsening in the horizontal direction only. Furthermore, due to the finite speed of sound in compressible models, it is sufficient to use a relatively small number of multigrid levels of $L\approx \log_2 \text{CFL}_{\text{h}}$ where $\text{CFL}_{\text{h}}=c_s\Delta t/\Delta x$ is the horizontal acoustic Courant number. The much higher vertical acoustic Courant number $\text{CFL}_{\text{v}}=c_s\Delta t/\Delta z$ does not cause any problems as vertical sound propagation is treated exactly by the line-relaxation smoother. Since advective transport is about an order of magnitude slower than acoustic pressure oscillations, $\text{CFL}_{\text{h}}\approx 10$ and $L\approx 4$ irrespective of the model resolution. As was demonstrated in \cite{Mueller2014,Sandbach2015}, this ``shallow'' multigrid works well, and avoids expensive global communications. It also significantly simplifies the parallel decomposition, since it is only necessary to (horizontally) partition the coarsest grid, which still has a large number of cells and allows a relative fine-grained domain decomposition. For example, one coarse grid cell could be assigned to a node on a supercomputer, exploiting additional shared-memory parallelism on the cells of the $4^{L-1}\approx 4^3=64$ fine grid cells.

The tensor-product multigrid algorithm was applied to more realistic model equations in \cite{Dedner2016} and its performance on a cluster with 16,384 GPUs was demonstrated in \cite{Mueller2015}.
\paragraph{Solvers for finite element discretisations.}
One challenge of standard latitude-longitude models, which is becoming more severe with increasing model resolution, is the convergence of grid-lines at the poles. Due to the resulting small grid cells at high latitudes this leads to severe timestep constrictions, slow solver convergence and poor parallel scalability due to global coupling at the poles. To overcome this problem, there has been a push towards using different meshes which avoid this issue (see review in \cite{Staniforth2012}). However, to ensure the accurate discretisation of the continuous equations and the exact conservation of certain physical quantities on those non-orthogonal grids requires advanced discretisations. While low-order finite-volume methods (\cite{Ringler2010,Thuburn2012,Thuburn2013}) and high-order collocated spectral element methods (\cite{Fournier2004,Giraldo2008}) exist, the mimetic finite-element approach developed in \cite{Cotter2012,Cotter2014,Thuburn2015} for the shallow water equations is particularly attractive since it generalises to arbitrary discretisation order, has good wave dispersion properties, avoids spurious computational modes and allows the approximate conservation of certain physical quantities in the discrete equations. At lowest order on orthogonal meshes it reduces to the extensively studied C-grid staggering in the horizontal direction.

This paper builds on the work in \cite{Melvin2018} which describes the recently developed Gungho dynamical core employed in the LFRic Model. The mimetic finite-element discretisation used there is combined with a vertical discretisation (described in \cite{Natale2016,Melvin2018a}), which is similar to Charney-Phillips staggering, and a mass-conserving finite-volume advection scheme.

A particular challenge of mimetic finite-element discretisations is the significantly more complex structure of the discretised linear equation system, which has to be solved repeatedly at every timestep. For traditional finite-difference and finite-volume discretisations on structured grids the Schur-complement can be formed, and -- provided the resulting pressure equation is solved to high-enough accuracy -- the preconditioner is exact. However, this is not possible for the finite element discretisations considered here since the velocity mass-matrix is not (block-) diagonal. Instead, the linear system is preconditioned by constructing an approximate Schur-complement using velocity mass-lumping. As demonstrated for a gravity-wave system in \cite{Mitchell2016}, this method is efficient if one V-cycle of the same bespoke tensor-product multigrid algorithm is used to solve the pressure system. As shown there, the method also works for next-to-lowest-order discretisations if a $p$- refinement is used on the finest level of the multigrid hierarchy.

In this paper it is shown how the method can be extended to solve the full equations of motion, the Euler equations for a perfect gas in a rotating frame. The efficiency of the multigrid algorithm is demonstrated by alternatively solving the pressure correction equation with a Krylov- subspace method. As will be shown by running on 100,000s of processing cores and solving problems with more than 1 billion ($10^9$) unknowns, multigrid also improves the parallel scalability since - in contrast to the Krylov- method - the multigrid V-cycle does not require any global reductions.
\paragraph{Implementation.}
To achieve optimal performance, an efficient implementation is required. In a continuously diversifying hardware landscape the code has to be performance portable. In general, the LFRic model uses an implementation which is based on the separation-of-concerns approach described in \cite{Adams2018}. The composability of iterative methods and preconditioners is exploited to easily swap components of the complex hierarchical solver in an object oriented Fortran 2003 framework (see Section 6 in \cite{Adams2018}).
\paragraph{Structure.}
This paper is organised as follows. After putting the research in context by reviewing related work in Section \ref{sec:literature_review}, the mixed finite element discretisation is described and the construction of a Schur-complement preconditioner for the linear system is discussed in Section \ref{sec:methods}. The properties of the elliptic pressure operator are used to construct a bespoke multigrid preconditioner. After outlining the parallel implementation in the LFRic framework in Section \ref{sec:implementation}, numerical results for performance and parallel scalability are presented in Section \ref{sec:results}. Conclusions are drawn and future work discussed in Section \ref{sec:conclusion}.
\section{Context and related work}\label{sec:literature_review}
\paragraph{Semi-implicit timestepping methods.}
One of the perceived drawbacks of semi-implicit models is the additional complexity required for solving a large non-linear problem. Iterative solvers introduce global communications, which potentially limits scalability and performance; this can become a serious issue for operational forecast systems which run on large supercomputers and have to deliver results on very tight timescales. Nevertheless, the comprehensive review of linear solvers techniques for atmospheric applications in \cite{Mueller2014} shows that semi-implicit models deserve serious consideration. Looking at actively developed dynamical cores which target massively parallel supercomputers, of the 11 non-hydrostatic implementations compared in the recent Dynamical Core Model Development Project (DCMIP16) presented in \cite{Ullrich2017}, two are semi-implicit: the Canadian GEM finite difference code (\cite{Yeh2002}) and FVM (\cite{Kuehnlein2019}), the next-generation finite volume version of the Integrated Forecasting Suite (IFS) (\cite{Temperton2001,Wedi2015}) developed at ECMWF; the current spectral-transform model used by ECMWF is also semi-implicit. To solve the linear equation system, those models use different approaches. IFS employs a global spectral transform, which diagonalises the operator in Fourier spaces. Although this approach inherently involves expensive all-to-all communications, scalability can be improved by exploiting properties of the spherical harmonics (\cite{Wedi2013}). The solution of the pressure system in IFS-FVM with a Generalised Conjugate Residual (GCR) method is described in \cite{Smolarkiewicz2011}; the preconditioner exactly inverts the vertical part of the operator, similar to the line relaxation strategy which is used in the smoother for the multigrid algorithm in this work. To solve the three dimensional elliptic boundary value problem in the GEM model, a vertical transform is used to reduce it to a set of decoupled two-dimensional problems, which are solved iteratively (see \cite{Cote1998}).

All of the above semi-implicit models use second-order accurate finite difference/volume discretisations or use the spectral-transform approach. In contrast, actively developed massively parallel high order spectral element codes include NUMA (\cite{Giraldo2013}), which uses an implicit-explicit (IMEX) time integrator, the CAM-SE/HOMME dynamical core \cite{Dennis2012} used in the ACME climate model and Tempest (\cite{Ullrich2014,Guerra2016}). Collocating the quadrature points with nodal points in the Continuous Galerkin (CG) formulation of NUMA results in a diagonal velocity mass matrix, which allows the construction of a Schur-complement pressure system. This system is then solved with an iterative method. This is in contrast to mixed finite element approach employed here, for which the velocity mass matrix is non-diagonal. To address this issue, the outer system is solved iteratively and preconditioned with an approximate Schur-complement based on a lumped mass matrix. It should be noted, however, that the construction of an efficient linear solver for the Discontinuous Galerkin (DG) version of NUMA is significantly more challenging since the numerical flux augments the velocity mass matrix by artificial diffusion terms; overcoming this problem is a topic of current research and it is argued in in \cite{Peraire2010,Kang2020} that hybridisable DG methods appear to be particularly suitable. As discussed below, applying a similar hybridised approach to the mixed finite element formulation is a promising direction for future work.
While the fully implicit version of NUMA has been optimised on modern chip architectures (see \cite{Abdi2017}), the massively parallel scaling tests in \cite{Mueller2015a} are reported for the horizontally explicit vertically implicit (HEVI) variant of the model, in which only the vertical couplings are treated implicitly. The same HEVI time integrator can also be used by the Tempest dynamical core. Again this vertically implicit solver is equivalent to the block-Jacobi smoother in the fully implicit multigrid algorithm and the preconditioner in \cite{Kuehnlein2019}.

The discretisation used by the semi-implicit GUSTO code developed at Imperial College is based on \cite{Natale2016,Yamazaki2017,Shipton2018} and is very similar to the one used in this work. In contrast to LFRic, which is developed for operational use, GUSTO is a research model implemented in the Firedrake Python code generation framework described in \cite{Rathgeber2017}. It uses the iterative solvers and preconditioners from the PETSc library (see \cite{Balay1997}) to solve the linear system. By default, the elliptic pressure operator is inverted with a black-box algebraic multigrid (AMG) algorithm. While in \cite{Mitchell2016} AMG has been shown to give comparable performance to the bespoke geometric multigrid preconditioners developed here, using off-the-shelf AMG libraries in the LFRic code is not feasible due to their incompatible parallelisation strategy. It would also introduce undesirable software dependencies for a key component of the model.
\paragraph{Parallel multigrid and atmospheric models.}
Multigrid algorithms allow the solution of ill-conditioned elliptic PDEs in a time which is proportional to the number of unknowns in the system. Due to this algorithmically optimal performance, they are often the method of choice for large scale applications in geophysical modelling. The hypre library (\cite{Falgout2002}) contains massively parallel multigrid implementations, including BoomerAMG, and has been shown to scale to 100,000s of cores in \cite{Baker2012}. Similarly, the scalability of the AMG solver in the DUNE library \cite{Blatt2006} has been demonstrated in \cite{Ippisch2011} and \cite{Notay2015} describes another highly parallel AMG implementation. In \cite{Gmeiner2014} massively parallel multigrid methods based on hybrid hierarchical grids are used to solve problems with $10^{12}$ unknowns on more than 200,000 compute cores.

While those results clearly show the significant potential of parallel multigrid algorithms, it is evident from the review in \cite{Mueller2014} that they are rarely used in semi-implicit atmospheric models. An exception is the recent implementation of the MPAS model. In \cite{Sandbach2015} it is shown that a semi-implicit methods with a multigrid solver can be competitive with fully explicit time integrators. A conditional semi-coarsening multigrid for the ENDGame dynamical core (\cite{Wood2014}) used by the Met Office is described in \cite{Buckeridge2010} and currently implemented in the Unified Model code. Another recent application of multigrid in a non-hydrostatic model is given in \cite{Yi2018}. Two-dimensional multigrid solvers for the Poisson-equation on different spherical grids relevant for atmospheric modelling are compared in \cite{Heikes2013}. More importantly, the work in \cite{Yang2016} showed that domain-decomposition based multigrid algorithms can be used to solve the Euler equations with $0.77\cdot 10^{12}$ unknowns.
This paper received the 2016 Gordon Prize for it showed that the code scales to 10 million cores and achieves 7.95 PetaFLOP performance on the TaihuLight supercomputer. Note, however, that none of the models described in this section is based on advanced finite element discretisations which are used in this work.
\section{Methods}\label{sec:methods}
In the following, the mimetic finite element discretisation of the Euler equations in the LFRic dynamical core is reviewed. By exploiting the structure of the pressure correction equation in the approximate Schur-complement solver, an efficient tensor-product multigrid algorithm is constructed.
\subsection{Continuous equations}
The dynamical core of the model solves the Euler equations for a perfect gas in a rotating frame
\begin{equation}
  \begin{aligned}
    \frac{\partial\vec{u}}{\partial t} & =  -\left(\nabla\times\vec{u}\right)\times\vec{u}-2\vec{\Omega}\times\vec{u}-\frac{1}{2}\nabla(\vec{u}\cdot\vec{u})-\nabla\Phi-c_{p}\theta\nabla\Pi,\\
    \frac{\partial\rho}{\partial t} & =  -\nabla\cdot\left(\rho\vec{u}\right),\\
\frac{\partial\theta}{\partial t} & =  -\vec{u}\cdot\nabla\theta,\\
\Pi^{\frac{1-\kappa}{\kappa}} &= \frac{R}{p_{0}}\rho\theta.
  \end{aligned}
  \label{eqn:continuous_equation}
\end{equation}
At every point in time the state of the atmosphere $\vec{x}=(\vec{u},\rho,\theta,\Pi)$ is described by the three dimensional fields for (vector-valued) velocity $\vec{u}$, density $\rho$, potential temperature $\theta$ and (Exner-) pressure $\Pi$. In Eq. \eqref{eqn:continuous_equation} $\Phi$ is the geopotential such that $\nabla\Phi=-\vec{g}$ where the vector $\vec{g}$ denotes the gravitational acceleration and the Earths rotation vector is denoted by $\vec{\Omega}$. $R$ is the gas constant per unit mass and $\kappa=R/c_p$, where $c_p$ is the specific heat at constant pressure; $p_0$ is a reference pressure. The equations are solved in a domain $\mathsf{D}$ which either describes the global atmosphere or a local area model (LAM); for further details on the relevant boundary conditions see \cite{Melvin2018}. While this paper describes the development of multigrid solvers for global models, the method can be easily adapted for LAMs.
\subsection{Finite element discretisation}
To discretise Eq. \eqref{eqn:continuous_equation} in space, the mimetic finite element discretisation from \cite{Cotter2012,Natale2016} is used. For this, four principal function spaces $\mathbb{W}_{i},\, i=0,1,2,3$, of varying degrees of continuity are constructed. Those function spaces are related by the de Rham complex (\cite{Bott2013})
\begin{equation}
\begin{array}{ccccccc}
  \mathbb{W}_{0} & \overset{\nabla}{\longrightarrow} &
  \mathbb{W}_{1} & \overset{\nabla\times}{\longrightarrow} &
  \mathbb{W}_{2} & \overset{\nabla\cdot}{\longrightarrow} &
  \mathbb{W}_{3}.
\end{array}\label{eqe:de_rham_complex}
\end{equation}
Pressure and density are naturally discretised in the entirely discontinuous space $\mathbb{W}_3$, while the space $\mathbb{W}_2$, which describes vector-valued fields with a continuous normal component, is used for velocity.
At order $p$ on hexahedral elements the space $\mathbb{W}_2$ is the Raviart-Thomas space $RT_{p}$ and $\mathbb{W}_3$ is the scalar Discontinuous Galerkin space $Q_{p}^{\text{DG}}$. As will be important later on, note that the space $\mathbb{W}_{2}=\mathbb{W}_{2}^{h}\oplus\mathbb{W}_{2}^{z}$ can be written as the direct sum of a component $\mathbb{W}_{2}^{z}$ which only contains vectors pointing in the vertical direction and the space $\mathbb{W}_2^{h}$ such that the elements of $\mathbb{W}_{2}^{h}$ are purely horizontal vector fields.
In the absence of orography those two spaces are orthogonal in the sense that
\begin{equation*}
  \int_{\mathsf{D}} \vec{u}^{(h)}\cdot \vec{u}^{(z)} \;dV = 0
  \qquad\text{for all $\vec{u}^{(h)}\in\mathbb{W}_{2}^{h}$ and $\vec{u}^{(z)}\in\mathbb{W}_{2}^{z}$.}
\end{equation*}
Note that $\mathbb{W}_{2}^{z}$ is continuous in the vertical direction and discontinuous in the tangential direction, whereas $\mathbb{W}_{2}^{h}$ is continuous in the horizontal direction only. To discretise the potential temperature field, an additional space $\mathbb{W}_{\theta}$ is introduced. $\mathbb{W}_{\theta}$ is the scalar-valued equivalent of $\mathbb{W}_{2}^{z}$ and has the same continuity.
\begin{figure}
  \begin{center}
    \includegraphics[width=0.5\linewidth]{\figdir/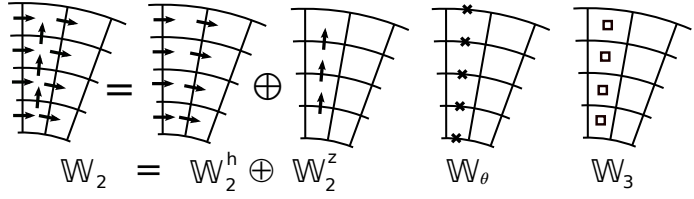}
    \caption{Function spaces used in the finite element discretisation}
    \label{fig:functionspaces}
  \end{center}
\end{figure}
The lowest order ($p=0$) function spaces are shown in Figure \ref{fig:functionspaces}.
Choosing suitable basis functions $\vec{v}_j(\vec{\chi})\in\mathbb{W}_2$, $\sigma_j(\vec{\chi})\in\mathbb{W}_3$ and $w_j(\vec{\chi})\in\mathbb{W}_\theta$ which depend on the spatial coordinate $\vec{\chi}$, the discretised fields at the $n$-th model timestep $t$ can be written as follows:
\begin{xalignat*}{2}
    \vec{u}^{n}(\vec{\chi}) &= \sum_j \widetilde{u}^n_j \vec{v}_j(\vec{\chi})\in\mathbb{W}_u,&
  \rho^n(\vec{\chi}) &=\sum_j \widetilde{\rho}^n_j \sigma_j(\vec{\chi}) \in\mathbb{W}_3,\\
  \theta^n(\vec{\chi}) &=\sum_j \widetilde{\theta}^n_j w_j(\vec{\chi})\in\mathbb{W}_{\theta},&
  \Pi^n(\vec{\chi})&=\sum_j \widetilde{\Pi}^n_j \sigma_j(\vec{\chi}) \in\mathbb{W}_3.
  \end{xalignat*}
For each quantity $a$ the corresponding vector of unknowns is written as $\widetilde{a}=[a_1,a_2,\dots]$.
\subsection{Linear system}\label{sec:linear_system}
The time discretisation described in \cite{Melvin2018} is semi-implicit and uses a finite-volume transport scheme. This requires the solution of a non-linear equation to obtain the state-vector $\widetilde{\vec{x}}^{n+1}=(\widetilde{u}^{n+1},\widetilde{\rho}^{n+1},\widetilde{\theta}^{n+1},\widetilde{\Pi}^{n+1})$ at the next timestep. The non-linear system can be written compactly as
\begin{equation}
  \mathcal{R}\left(\widetilde{\vec{x}}^{n+1}\right)=0.\label{eqn:nonlinear_system}
\end{equation}
Eq. \eqref{eqn:nonlinear_system} is solved iteratively with a quasi-Newton method. For this a sequence of states $\widetilde{\vec{x}}^{(k)}$, $k=0,1,2,\dots,N_{\text{NL}}$ with $\widetilde{\vec{x}}^{(0)}=\widetilde{\vec{x}}^{n}$, $\widetilde{\vec{x}}^{(N_{\text{NL}})}=\widetilde{\vec{x}}^{(n+1)}$ is constructed such that
\begin{equation}
  \mathcal{L}(\widetilde{\vec{x}}^*)\widetilde{\vec{x}}^{\prime} = -\mathcal{R}\left(\widetilde{\vec{x}}^{(k)}\right)\qquad\text{with $\widetilde{\vec{x}}^{\prime}=\widetilde{\vec{x}}^{(k+1)}-\widetilde{\vec{x}}^{(k)}$}.
  \label{eqn:QN_linear_system}
\end{equation}
The linear operator $\mathcal{L}(\widetilde{\vec{x}}^*)$, which needs to be inverted in every Newton step, is an approximation to the Jacobian of $\mathcal{R}$. Following \cite{Wood2014}, it is obtained by linearisation around a reference state $\widetilde{\vec{x}}^*=(0,\widetilde{\rho}^*,\widetilde{\theta}^*,\widetilde{\Pi}^*)$, which is updated at every timestep.
Introducing $\widetilde{\vec{x}}^{\prime}=(\widetilde{u}^{\prime},\widetilde{\rho}^{\prime},\widetilde{\theta}^{\prime},\widetilde{\Pi}^{\prime})$ and following \cite{Melvin2018}, the linear system in Eq. \eqref{eqn:QN_linear_system} can be written down in matrix form as
\begin{equation}
  \begin{pmatrix}
    M_{2}^{\mu,C}
    &
    & - P_{2\theta}^{\Pi^{*}}
    & - G^{\theta^{*}}
    \\
    D^{\rho^{*}}
    & M_{3}
    &
    &
    \\
    P_{\theta2}^{\theta^{*}}
    &
    & M_{\theta}
    &
    \\
    &-M_{3}^{\rho^{*}}
    &- P_{3\theta}^{*}
    &M_{3}^{\Pi^{*}}
  \end{pmatrix}
  \begin{pmatrix}
    \widetilde{u}^{\prime}\\
    \widetilde{\rho}^{\prime}\\
    \widetilde{\theta}^{\prime}\\
    \widetilde{\Pi}^{\prime}
  \end{pmatrix}
  =
  \begin{pmatrix}
    -{\cal R}_{u}\\
    -{\cal R}_{\rho}\\
    -{\cal R}_{\theta}\\
    -{\cal R}_{\Pi}
  \end{pmatrix}.
  \label{eqn:linear_matrix}
\end{equation}
The exact form of the individual operators in Eq. \eqref{eqn:linear_matrix} is given in \cite{Melvin2018} and the matrix $D^{\rho^{*}}$ is defined as $D^{\rho^{*}}\widetilde{u} := D\left(\widetilde{f}^*\right)\equiv D\left(\widetilde{\rho}^{*}\widetilde{u}\right)$ where the mass flux $f^*$ is defined as the product of the reference density $\rho^*$ sampled at velocity nodal points pointwise multiplied by the velocity field $\widetilde{u}$. Note, that the expression in \cite{Melvin2018} for ${\cal R}_{\Pi}$, their (81), is incorrect and it should be
\begin{equation}
  {\cal R}_{\Pi} \equiv \left\langle \widehat{\sigma},\det\mathbf{J}\left[ 1 - \frac{p_0}{R}\frac{\left(\widehat{\Pi}^{\left(k\right)}\right)^{\frac{1-\kappa}{\kappa}}}{\widehat{\rho}^{\left(k\right)}\widehat{\theta}^{\left(k\right)}} \right] \right\rangle,\label{eqn:Rhs-pi}
\end{equation}
such that both the linearised left hand side and nonlinear right hand side are non-dimensionalised.
To interpret the different operators, it is instructive to also write down the continuum equivalent of the equations for the state $(\vec{u}',\rho',\theta',\Pi')$
\begin{equation}
  \begin{aligned}
    \vec{u}^{\prime}+\tau_u\Delta t\left(\mu\hat{\mathbf{z}}(\hat{\mathbf{z}}\cdot \vec{u}^{\prime}) + 2\bm{\Omega}\times \vec{u}^{\prime} \right) +\tau_u\Delta tc_p \left(\theta^\prime\nabla\Pi^*+\theta^*\nabla\Pi^\prime\right) &= \vec{r}_u, \\
    \rho^\prime + \tau_\rho \Delta t\nabla\cdot \left(\rho^*\vec{u}^\prime\right) &= r_\rho,\\
    \theta^\prime + \tau_\theta \Delta t \vec{u}^\prime\cdot\nabla \theta^* &= r_\theta,\\
    \frac{\Pi^\prime}{\Pi^*} - \frac{\kappa}{1-\kappa}\left(\frac{\rho^\prime}{\rho^*}+\frac{\theta^\prime}{\theta^*}\right) &= r_\Pi.
  \end{aligned}
  \label{eqn:linear_continuum}
\end{equation}
where $\tau_{u,\rho,\theta}=\frac{1}{2}$ are relaxation parameters. $\rho^*$, $\theta^*$ and $\Pi^*$ are the continuous reference profiles around which the equation is linearised. The unit normal vector in the vertical direction is denoted as $\hat{\mathbf{z}}$ and the quantity $\mu$ is defined as $\mu=1+\tau_u \tau_\theta \Delta t^2 N^2$ with the Brunt-V\"{a}is\"{a}l\"{a} frequency $N^2=g(\partial_z\theta^*)/\theta^*$. In contrast to \cite{Wood2014}, the horizontal couplings are not neglected in $P_{\theta 2}^{\theta^*}$, and will only be dropped in the approximate Schur-complement constructed in Section \ref{sec:schur_complement}. The block-diagonal entries in the $4\times 4$ matrix in Eq. \eqref{eqn:linear_matrix} are modified mass matrices of the $\mathbb{W}_{2}$, $\mathbb{W}_3$ and $\mathbb{W}_{\theta}$ spaces, possibly weighted by reference profiles ($M_3^{\Pi^*}$, similar to the off-diagonal $M_3^{\rho^*}$). The term in the upper left corner of the matrix in Eq. \eqref{eqn:linear_matrix} is the velocity mass matrix augmented by contributions from Rayleigh damping (optionally only applied to the vertical component of the velocity vector near the model lid, see \cite{Melvin2018}) and the implicit treatment of the Coriolis term,
\begin{equation}
  M_2^{\mu,C} = M_2 +\tau_u \Delta t (M_\mu + M_C),\label{eqn:M2muCdef}
\end{equation}
where
\begin{equation*}
  \left(M_C\right)_{ij}=2\int_\mathsf{D}\vec{v}_i\cdot (\Omega\times\vec{v}_j)\;dV.
\end{equation*}
While $M_3$, $M_3^{\Pi^*}$ and $M_3^{\rho^*}$ do not contain couplings to unknowns in neighbouring cells, and $M_\theta$ only couples between unknowns in the same vertical column, $M_2^{\mu,C}$ contains couplings in all directions. This prevents the exact solution of Eq. \eqref{eqn:linear_matrix} with a Schur-complement approach as in \cite{Wood2014} since the inverse of the $M_2^{\mu,C}$ is dense.
Instead, the system in Eq. \eqref{eqn:linear_matrix} is solved with an iterative Krylov subspace solver, which only requires application of the sparse operator $\mathcal{L}(\widetilde{\vec{x}}^*)$ itself. The solver is preconditioned with the approximate Schur complement described in the following section.
\subsection{Schur complement preconditioner}\label{sec:schur_complement}
To obtain an approximate solution of the linear system in Eq. \eqref{eqn:linear_matrix}, first all instances of the mass matrix $M_\theta$ are replaced by a lumped, diagonal version $\mathring{M}_{\theta}$, such that the diagonal entries of $\mathring{M}_{\theta}$ are the row-sums of $M_{\theta}$.
As in \cite{Wood2014} only the part of $P_{\theta 2}^{\theta^*}$ which acts on the vertical part of the velocity field is kept. The resulting operator $P_{\theta 2}^{\theta^*,z}$ maps from the subspace $\mathbb{W}_2^z\subset \mathbb{W}_2$ to $\mathbb{W}_{\theta}$.

Algebraically, the following steps correspond to multiplication by the upper block-triangular matrix (Step 1), solution of the block-diagonal matrix (Step 2) and back-substitution through multiplication by the lower block-triangular matrix (Step 3) in the Schur-complement approach (\cite{Zhang2006}).
\paragraph{Step 1a:} Use
\begin{equation}
  \widetilde{\theta}^{\prime} = \mathring{M}_\theta^{-1}\left(-P_{\theta2}^{\theta^*,z}\widetilde{u}^{\prime}-\mathcal{R}_\theta\right),\label{eqn:eliminate_theta}
\end{equation}
to eliminate $\theta$. In the resulting $3\times 3$ system, replace the matrix $\overline{M}_2^{\mu,C}:=M_2^{\mu,C}+P_{2\theta}^{\Pi^*}\mathring{M}_{\theta}^{-1}P_{\theta2}^{\theta^*,z}$ on the diagonal by a lumped diagonal approximation $\mathring{M}_{2}^{\mu,C}$, such that the diagonal entries of $\mathring{M}_2^{\mu,C}$ are the row-sums of $\overline{M}_2^{\mu,C}$. This leads to a system for $\widetilde{u}^{\prime}$, $\widetilde{\rho}^{\prime}$ and $\widetilde{\Pi}^{\prime}$ only:
\begin{equation}
  \begin{pmatrix}
    \mathring{M}_2^{\mu,C} & & -G^{\theta^*}\\
    D^{\rho^*} & M_3 & \\
    P_{3\theta}^* \mathring{M}_{\theta}^{-1}P_{\theta2}^{\theta^*,z} & -M_3^{\rho^*} & M_3^{\Pi^*}
  \end{pmatrix}
    \begin{pmatrix}
    \widetilde{u}^{\prime}\\
    \widetilde{\rho}^{\prime}\\
    \widetilde{\Pi}^{\prime}
    \end{pmatrix}
  =
  \begin{pmatrix}
    -\overline{{\cal R}}_{\mathbf{u}}\\
    -{\cal R}_{\rho}\\
    -\overline{{\cal R}}_{\Pi}
  \end{pmatrix},
  \label{eqn:linear_system3x3}
\end{equation}
with
\begin{xalignat}{2}
\overline{{\cal R}}_{\mathbf{u}} &= {\cal R}_{u}-P_{2\theta}^{\Pi^{*}}\mathring{M}_{\theta}^{-1}{\cal R}_{\theta}, &
\overline{\cal R}_{\Pi} &=
{\cal R}_{\Pi}-P_{3\theta}^{*}\mathring{M}_{\theta}^{-1}{\cal R}_{\theta}.
\end{xalignat}
Note that - in contrast to Eq. \eqref{eqn:linear_matrix} - the (block-) diagonal entries of the $3\times 3$ system in Eq. \eqref{eqn:linear_system3x3} have sparse inverses, and it is possible to form the exact Schur complement.
\paragraph{Step 1b:} Similarly, eliminate density from Eq. \eqref{eqn:linear_system3x3} using
\begin{equation}
  \widetilde{\rho}^{\prime} = M_3^{-1} \left(-D^{\rho^*}\widetilde{u}^{\prime}-{\cal R}_{\rho}\right),\label{eqn:eliminate_rho}
\end{equation}
to obtain a $2\times 2$ system for $\widetilde{u}^{\prime}$, $\widetilde{\Pi}^{\prime}$. Finally, eliminate velocity with
\begin{equation}
  \widetilde{u}^{\prime} = \left(\mathring{M}_2^{\mu,C}\right)^{-1} \left(G^{\theta^*}\widetilde{\Pi}^{\prime}-\overline{\cal R}_u\right),\label{eqn:eliminate_u}
\end{equation}
to get an equation for the pressure increment only,
\begin{equation}
  H\widetilde{\Pi}^{\prime} = {\cal B}=-\overline{\cal R}_{\Pi} + \left(\mathring{M}_2^{\mu,C}\right)^{-1}\overline{\cal R}_u - M_3^{\rho^*}M_3^{-1} {\cal R}_{\rho}.
  \label{eqn:helmholtz}
\end{equation}
The Helmholtz operator $H:\mathbb{W}_3\rightarrow \mathbb{W}_3$ is defined as
\begin{equation}
  H = M_3^{\Pi^*} + \left(P_{3\theta}^{*}\mathring{M}_{\theta}^{-1}P_{\theta2}^{\theta^*,z}+M_3^{\rho^*}M_3^{-1}D^{\rho^*}\right)\left(\mathring{M}_2^{\mu,C}\right)^{-1}G^{\theta^*}.\label{eqn:helmholtz_operator}
\end{equation}
\paragraph{Step 2:} Approximately solve the Helmholtz equation in Eq. \eqref{eqn:helmholtz} for $\widetilde{\Pi}^{\prime}$ . For this one multigrid V-cycle as described in Section \ref{sec:multigrid} is used.
\paragraph{Step 3:} Given $\widetilde{\Pi}^{\prime}$, recover $\widetilde{u}^{\prime}$, $\widetilde{\rho}^{\prime}$ and $\widetilde{\theta}^{\prime}$ using Eqs. \eqref{eqn:eliminate_u}, \eqref{eqn:eliminate_rho} and \eqref{eqn:eliminate_theta}.
\subsection{Structure of the Helmholtz operator}\label{eqn:helmholtz_structure}
Understanding the structure of the Helmholtz operator $H$ is crucial for the construction of a robust multigrid algorithm for the approximate solution of Eq. \eqref{eqn:helmholtz}. The tensor-product multigrid method which will be used here was first described for simpler equations and discretisations in \cite{Borm2001} and applied to mixed finite element problems in atmospheric modelling in \cite{Mitchell2016}.

\begin{figure}
  \begin{center}
    \includegraphics[width=0.25\linewidth]{\figdir/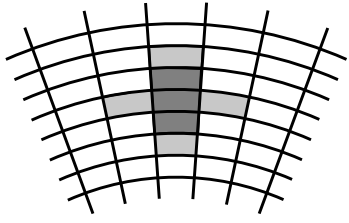}
    \caption{Stencil of the Helmholtz operator $H$ (all gray cells) and of the operator $\widehat{H}_z$ (dark gray cells). For clarity, only a two-dimensional cross section of the stencil is shown, in three dimensions $H$ has a nine entries (instead of seven as for the standard near-neighbour stencil) and $H_z$ has three entries.}
    \label{fig:helmholtz_stencil}
  \end{center}
\end{figure}

First consider the sparsity pattern of the Helmholtz operator $H$. In each cell of the grid, it contains couplings to its four direct horizontal neighbours. In addition, it couples to the two cells immediately above and the two cells immediately below. Including the self-coupling, this results in a 9-cell stencil, independent of the order of discretisation $p$ (see Figure \ref{fig:helmholtz_stencil}).
Secondly, it is important to take into account how the components of $H$ depend on the timestep size $\Delta t$, the horizontal grid spacing $\Delta x$ and the vertical grid spacing $\Delta z$. For this, first note that the weighted weak derivative $D^{\rho^*}$ can be decomposed into a vertical and horizontal part $D^{\rho^*}=D^{\rho^*,z}+D^{\rho^*,h}$ with $D^{\rho^*,z}:\mathbb{W}_2^z\rightarrow \mathbb{W}_3$ and $D^{\rho^*,h}:\mathbb{W}_2^h\rightarrow \mathbb{W}_3$. Since the lumped mass matrix is diagonal, it is the sum of two terms, $\mathring{M}_2^{\mu,C}=\mathring{M}_2^{\mu,C,z}+\mathring{M}_2^{\mu,C,h}$ with $\mathring{M}_2^{\mu,C,h}:\mathbb{W}_2^h\rightarrow\mathbb{W}_2^h$ and $\mathring{M}_2^{\mu,C,z}:\mathbb{W}_2^z\rightarrow\mathbb{W}_2^z$. Using this decomposition, the Helmholtz operator can be written as the sum of four terms
\begin{equation}
  \begin{aligned}
    H &= \underbrace{M_3^{\Pi^*}}_{H_0} + \underbrace{P_{3\theta}^{*}\mathring{M}_{\theta}^{-1}P_{\theta2}^{\theta^*,z}\left(\mathring{M}_2^{\mu,C,z}\right)^{-1}G^{\theta^*}}_{D^z_{1}}+ \underbrace{M_3^{\rho^*}M_3^{-1}D^{\rho^*,z}\left(\mathring{M}_2^{\mu,C,z}\right)^{-1}G^{\theta^*}}_{D^z_2}+    \underbrace{M_3^{\rho^*}M_3^{-1}D^{\rho^*,h}\left(\mathring{M}_2^{\mu,C,h}\right)^{-1}G^{\theta^*}}_{D^h_2}.
  \end{aligned}\label{eqn:Helmholtz_operator}
\end{equation}
In order to interpret the different parts of $H$ it is constructive to derive the corresponding Schur-complement operator for the continuous linear system which is given in Eq. \eqref{eqn:linear_continuum}. To be consistent with the substitution $P_{\theta2}^{\rho^*}\mapsto P_{\theta2}^{\rho^*,z}$ above, the third equation of Eq. \eqref{eqn:linear_continuum} is replaced by $\theta^\prime + \tau_\theta \Delta t u_z^\prime \partial_z \theta^* = r_\theta$. The resulting pressure operator is
\begin{equation}
  \begin{aligned}
  h &= \underbrace{\frac{1}{\Pi^*}}_{h_0}
    \;\;\underbrace{-\tau_\rho\tau_u \Delta t^2 c_p\frac{\kappa}{1-\kappa} \mu^{-1} (\partial_z\theta^*)\partial_z}_{d_1^z}
    \underbrace{-\;\; \tau_\rho \tau_u \Delta t^2 c_p\frac{\kappa}{1-\kappa}\frac{\partial_z\left(\mu^{-1} \rho^*\theta^*\partial_z \right)}{\rho^*}}_{d_2^z}
  \underbrace{- \tau_\rho \tau_u \Delta t^2 c_p\frac{\kappa}{1-\kappa}\frac{\nabla_h\cdot\left(\rho^*\theta^*\nabla_h\right)}{\rho^*}}_{d_2^h},
  \end{aligned}
  \label{eqn:continuum_pressure_operator}
\end{equation}
where $\nabla_h$ is the horizontal gradient operator. Up to scaling by the local volume of the grid cell it is therefore possible to identify $H_0\doteq h_0$, $D_1^z\doteq d_1^z$, $D_2^z\doteq d_2^z$ and $D_2^h\doteq d_2^h$. With the definitions of the linear operators in \cite{Melvin2018}, it is easy to see that the different parts of $H$ depend on the grid spacing and timestep size as
\begin{xalignat}{3}
  H_0 &\propto \Delta x^2\cdot \Delta z, &
  D_1^z,D_2^z&\propto \frac{\Delta x^2\cdot \Delta t^2}{\Delta z}, &
  D_2^h&\propto \Delta z\cdot \Delta t^2
.\label{eqn:DH_ratios}
\end{xalignat}
Further observe that with $T^*=\Pi^*\theta^*$ and the ratio of the specific heat capacities $\kappa/(1-\kappa)=c_p/c_v$ the (squared) speed of sound is given by $c_s^2=c_p/c_v RT^*=c_p\frac{\kappa}{1-\kappa}\Pi^*\theta^*$. Assuming that the reference profiles are slowly varying, this implies that the ratios
$d_1^z:h_0$, $d_2^z:h_0$ and $d_2^h:h_0$ scale as
\begin{xalignat}{3}
  d_1^z:h_0 &\propto \frac{N^2c_s^2}{g}\Delta t^2,&
  d_2^z:h_0 &\propto c_s^2\Delta t^2,&
  d_2^h:h_0 &\propto c_s^2\Delta t^2.
\end{xalignat}
Combining this with Eq. \eqref{eqn:DH_ratios} results in the estimates
\begin{xalignat*}{3}
  D_1^z:H_0 &\propto \frac{N^2 c_s^2}{g}\cdot \frac{\Delta t^2}{\Delta z}, &
  D_2^z:H_0 &\propto \left(\frac{c_s\Delta t}{\Delta z}\right)^2=\text{CFL}_{\text{v}}^2,&
  D_2^h:H_0 &\propto \left(\frac{c_s\Delta t}{\Delta x}\right)^2=\text{CFL}_{\text{h}}^2,
\end{xalignat*}
where $\text{CFL}_{\text{v}}$ and $\text{CFL}_{\text{h}}$ are the vertical and horizontal acoustic Courant numbers. Note also that the relative size of $D_2^z$ and $D_2^h$ is given by the squared aspect ratio $(\Delta x/\Delta z)^2$ and the relative size of $D_1^z$ and $D_2^z$ decreases $\propto \Delta z$ as the vertical grid spacing goes to zero.

To proceed further, the Helmholtz operator is split into two parts, $H=H_z+\delta H$ such that $H_z$ contains the couplings to neighbouring cells in the vertical direction only. If the degrees of freedom are ordered consecutively in the vertical direction, $H_z$ is a block-diagonal matrix. Each block describes the couplings in one vertical column; furthermore, solution of the system $H_z\widetilde{\Pi}=r_{\Pi}$ for some right hand side $r_\Pi$ requires the independent solution of a block-pentadiagonal system in each column. Following the scaling arguments above and observing that the operators $D_1^z$ and $D_2^z$ only contribute to $H_z$, it can be seen that for high aspect ratios $\Delta z\ll \Delta x$ the dominant part of the operator $H$ is given by $H_z$. This observation is crucial for the following construction of a robust tensor-product multigrid algorithm.
\subsection{Multigrid}\label{sec:multigrid}
Starting from some initial guess, an approximate solution of Eq. \eqref{eqn:helmholtz} can be obtained with a block-Jacobi-iteration. To avoid the expensive block-pentadiagonal solve, the next-to-nearest neighbour couplings in $H_z$ are dropped to obtain a block-tridiagonal $\widehat{H}_z$, see Figure \ref{fig:helmholtz_stencil}. With this matrix, one iteration of the block-Jacobi method is
\begin{equation}
\widetilde{\Pi}^\prime \mapsfrom \widetilde{\Pi}^{\prime} + \omega \widehat{H}_z^{-1}\left({\cal B}-H\widetilde{\Pi}^{\prime}\right)\label{eqn:jacobi},
\end{equation}
where $\omega$ is an over-relaxation factor. The shorthand $\widetilde{\Pi}^{\prime}\mapsfrom \textsf{BlockJacobi}(H,\mathcal{B},\widetilde{\Pi}^{\prime},\omega,n_{\text{Jac}})$ is used for $n_{\text{Jac}}$ applications of the block- Jacobi iteration in Eq. \eqref{eqn:jacobi}. Multiplication by $\widehat{H}_z^{-1}$ in Eq. \eqref{eqn:jacobi} corresponds to the solution of block-tridiagonal linear system, which can be carried out independently in each vertical column. The tridiagonal solve can be done, for example, with the Thomas algorithm (see e.g. \cite{Press2007}). When applying $H$ to $\widetilde{\Pi}^{\prime}$ to calculate the residual ${\cal B}-H\widetilde{\Pi}^{\prime}$ in Eq. \eqref{eqn:jacobi}, the vertical terms $D_2^z$ and $D_1^z$ in Eq. \eqref{eqn:Helmholtz_operator} are treated exactly.

It is well known that stationary methods such as the Jacobi iteration converge extremely slowly since they only reduce the high-frequency error components. This issue is overcome by multigrid methods (\cite{Trottenberg2000} contains a comprehensive treatment of the topic), which construct a hierarchy of grids with associated (nested) finite element function spaces, in particular $\mathbb{W}_3=\mathbb{W}_3^{(1)}\supset \mathbb{W}_3^{(2)}\supset \dots\supset\mathbb{W}_3^{(L)}$. Following the tensor-product approach in \cite{Borm2001}, the grid is only coarsened in the horizontal direction. By applying a small number of smoother iterations on each level, the error is reduced at all length scales. In the following the index $\ell\in\{1,\dots,L\}$ is used to label the multigrid level, with $L=1$ corresponding to the fine grid level on which the solution is to be found. Let $\{\widetilde{\Pi}^{\prime(\ell)}\}$ and $\{{\cal B}^{(\ell)}\}$ be the set of solution vectors and right hand sides on all levels, with $\widetilde{\Pi}^{\prime(1)}=\widetilde{\Pi}^\prime$ and ${\cal B}^{(1)}={\cal B}$. Since the function spaces are nested, the obvious prolongation $\mathfrak{P}:\mathbb{W}_3^{(\ell+1)}\rightarrow\mathbb{W}_3^{(\ell)}$ from a coarse space to the next-finer multigrid level is the natural injection:
\begin{equation}
  \mathfrak{P}: \Pi^{\prime(\ell+1)} \mapsto \Pi_{\mathfrak{P}}^{\prime(\ell)},\label{eqn:prolongation}
\end{equation}
with
\begin{equation*}
  \Pi_{\mathfrak{P}}^{\prime(\ell)}(\vec{\chi}) = \Pi^{\prime(\ell+1)}(\vec{\chi})\qquad\text{for all points $\vec{\chi}\in\textsf{D}$.}
\end{equation*}
The corresponding linear operator acting on the dof-vector $\widetilde{\Pi}^{\prime(\ell+1)}$ can be written as $\textsf{Prolongate}\left(\widetilde{\Pi}^{\prime(\ell+1)}\right)$. Eq. \eqref{eqn:prolongation} naturally induces a restriction $\mathfrak{R}:\mathbb{W}_3^{(\ell)*}\rightarrow\mathbb{W}_3^{(\ell+1)*}$ on the corresponding dual spaces (denoted by an asterisk $*$):
\begin{equation*}
  \mathfrak{R}: r^{(\ell)} \mapsto r^{(\ell+1)},
\end{equation*}
with
\begin{equation*}
  r^{(\ell+1)}\left(\Pi^{\prime(\ell+1)}\right)=r^{(\ell)}\left(\mathfrak{P}\left(\Pi^{\prime(\ell+1)}\right)\right),
\end{equation*}
for all functions $\Pi^{\prime(\ell+1)}\in\mathbb{W}_3^{(\ell+1)}$.
The corresponding linear operator acting on the vector ${\cal R}^{(\ell)}$ representing the dual 1-form $r^{(\ell)}$ is written as $\textsf{Restrict}\left({\cal R}^{(\ell)}\right)$; note that the level-dependent residual ${\cal R}^{(\ell)}$ is different from the quantities that appear on the right hand side of \eqref{eqn:linear_matrix}. The Helmholtz-operators on the coarse levels are constructed by representing the reference profiles on those levels and re-discretising the operator. This is more efficient than assembling it via the expensive Galerkin triple matrix-product.

Based on those ingredients, it is now possible to write down the recursive multigrid V-cycle in Algorithm \ref{alg:mg_vcycle}. Starting from some initial guess $\widetilde{\Pi}^\prime = \widetilde{\Pi}^{\prime(1)}$ and right hand side ${\cal B}^{(1)}={\cal B}$ on the finest level, this reduces the error by recursively solving the residual equation on the next-coarsest level.
\begin{algorithm}
    \caption{Multigrid V-cycle, $\Sigma^{\ell}=\{H^{(\ell)},\widetilde{\Pi}^{\prime(\ell)},{\cal B}^{(\ell)},{\cal R}^{(\ell)}\}$ $\textsf{MGVcycle}(\Sigma^{(\ell)},\ell,\omega,n_{\text{pre}},n_{\text{post}})$}
  \label{alg:mg_vcycle}
  \begin{algorithmic}[1]
  \IF{$\ell=L$}
  \STATE{$\widetilde{\Pi}^{\prime(L)}\mapsfrom \textsf{CoarseSolve}(H^{(L)},{\cal B}^{(L)})$}
  \ELSE
  \STATE{$\widetilde{\Pi}^{\prime(\ell)}\mapsfrom \textsf{BlockJacobi}(H^{(\ell)},\mathcal{B}^{(\ell)},\widetilde{\Pi}^{\prime(\ell)},\omega,n_{\text{pre}})$}
  \STATE{${\cal R}^{(\ell)}\mapsfrom {\cal B}^{(\ell)}-H^{(\ell)}\widetilde{\Pi}^{\prime(\ell)}$}
  \STATE{${\cal B}^{(\ell+1)}\mapsfrom \textsf{Restrict}({\cal R}^{(\ell)})$}
  \STATE{$\widetilde{\Pi}^{\prime(\ell+1)}\mapsfrom 0$}
  \STATE{$\widetilde{\Pi}^{\prime(\ell+1)}\mapsfrom\textsf{MGVcycle}(\{\Sigma^{(\ell)}\},\ell+1,\omega,n_{\text{pre}},n_{\text{post}})$}
  \STATE{$\widetilde{\Pi}^{\prime(\ell)}\mapsfrom\widetilde{\Pi}^{\prime(\ell)}+\textsf{Prolongate}(\widetilde{\Pi}^{\prime(\ell+1)})$}
  \STATE{$\widetilde{\Pi}^{\prime(\ell)}\mapsfrom \textsf{BlockJacobi}(H^{(\ell)},\mathcal{B}^{(\ell)},\widetilde{\Pi}^{\prime(\ell)},\omega,n_{\text{post}})$}
  \ENDIF
  \end{algorithmic}
\end{algorithm}
The natural way of (approximately) solving the equation on the coarsest level $\ell=L$ is by inverting the matrix directly or applying some Krylov-subspace method. In a parallel implementation this requires expensive global communications. As explained in \cite{Sandbach2015,Mueller2014} this is not necessary in our case. To see this, observe that the relative size of the zero order term $H_0$ and the second derivative in the horizontal direction $D_2^h$ is proportional to the squared, inverse grid spacing $\Delta x$. Since the grid spacing doubles on each subsequent level, the relative size of the two terms in the Helmholtz operator reduces to \mbox{$4^{-(\ell-1)}\text{CFL}_{\text{h}}^2$} where, as above, $\text{CFL}_{\text{h}}$ is the acoustic Courant number in the horizontal direction. As the vertical terms are treated exactly in the block-Jacobi smoother, the condition number of the Helmholtz operator will be $\mathcal{O}(1)$ on levels $\ell\gtrsim \log_2 (\text{CFL}_{\text{h}})+1$. Hence, it is sufficient to pick $L=\log_2(\text{CFL}_{\text{h}})+1$ and simply apply a few iterations of the block-Jacobi smoother. For typical atmospheric applications, $\text{CFL}_{\text{h}} \approx 10$, and hence it is sufficient to work with $L\approx 4$ levels. As demonstrated in \cite{Mueller2014} this shallow multigrid approach also greatly reduces global communications.

\subsection{Computational complexity}
Although the multigrid method requires additional calculations on the coarse levels, its computational complexity is proportional to the number of unknowns. The time spent in one multigrid V-cycle in Alg. \ref{alg:mg_vcycle} is dominated by two contributions: the multiplication with the matrix $H^{(\ell)}$ and the vertical solve, i.e. the application of $(\widehat{H}_z^{(\ell)})^{-1}$. Those operations are required in the residual calculation and the block-Jacobi algorithm, which is used for both for pre-/post-smoothing and for the approximate solution of the coarse level system with $n_{\text{coarse}}$ block-Jacobi iterations. Let $C_{H}$ and $C_{\widehat{H}_z}$ be the cost per unknown for those two operations. For a pressure system with $N$ unknowns the computational cost per multigrid V-cycle is
\begin{equation*}
  \begin{aligned}
    \text{Cost}_{\text{V-cycle}} &= \left((n_{\text{pre}}+n_{\text{post}})(C_H+C_{\widehat{H}_z})+C_H\right)N\sum_{\ell=1}^{L-1} 4^{-\ell+1}+ n_{\text{coarse}} (C_H+H_{\widehat{H}_z})N4^{-L}\\
    &\approx \frac{4}{3}\left((n_{\text{pre}}+n_{\text{post}})(C_H+C_{\widehat{H}_z})+C_H\right)N,
  \end{aligned}\label{eqn:MG_cost_estimate}
\end{equation*}
where the approximation in the last line is valid for $4^{-L+1}\ll 1$.

In contrast, solving the Helmholtz pressure system with $n_{\text{iter}}$ iterations of a BiCGStab method, preconditioned by $\widehat{H}_z^{-1}$ involves a cost of
\begin{equation*}
  \text{Cost}_{\text{BiCGStab}} = \left(2n_{\text{iter}}(C_H+C_{\widehat{H}_z})+C_H\right)N.
\end{equation*}
While the multigrid V-cycle contains more nearest-neighbour parallel communication in the form of halo exchanges, those can easily be overlapped with computations via asynchronous MPI calls.
In contrast, the BiCGStab solver requires 4 global sums per iteration (including one for monitoring convergence in the residual norm). While communication avoiding variants of Krylov solvers exist (see the overview in \cite{Hoemmen2010} and recent work on BiCGStab in \cite{Carson2013}), this will not overcome the fundamental issue. Several BiCGStab iterations with global communications are required to achieve the same reduction in the pressure residual as in a multigrid V-cycle. As the numerical results in Section \ref{sec:results_parallel} demonstrate, this reduces the scalability of Krylov-subspace solvers for the pressure correction equation.
\subsection{Memory requirements}
The memory requirements of the different solvers for the pressure equation are quantified by counting the number of dof-vectors they need to store. Since the matrix $H$ has a 9-point stencil and the LU-factorisation of the tridiagonal matrix $\widehat{H}_z$ is required in the block-Jacobi iteration in Eq. \eqref{eqn:jacobi}, storing the matrix requires the same memory as 12 dof-vectors. As can be seen from Alg. \ref{alg:mg_vcycle}, on each level of the multigrid hierarchy the three vectors $\widetilde{\Pi}^{(\ell)}$, $\mathcal{B}^{(\ell)}$ and $\mathcal{R}^{(\ell)}$ are stored in addition to the Helmholtz matrix, resulting in a total memory requirement of 15 dof-vectors on each level. However, since the number of unknowns is reduced by a factor of four in each coarsening step, the memory requirements on the coarser levels are significantly reduced. In the standalone multigrid iteration two additional vectors are required on the finest level to monitor convergence. Assuming that a dof-vector on the finest level contains $N$ unknowns, this results in a total memory requirement of
\begin{equation*}
  \begin{aligned}
    \text{Memory}_{\text{Multigrid}} &= 15\left(1+\frac{1}{4}+\frac{1}{16}+\dots\right)N+2N < 22 N
    \end{aligned}
\end{equation*}
for the multigrid method. This should be compared to the BiCGStab solver: in addition to the solution, right-hand-side and the matrix, this uses eight temporary vectors, resulting in a total storage requirement of
\begin{equation*}
  \text{Memory}_{\text{BiCGStab}} = 22N.
\end{equation*}
The memory requirements of other solver combinations considered in this work are in the same ballpark. Using a standalone block-Jacobi iteration requires the storage of 16 dof-vectors, whereas the equivalent of no more than 28 dof-vectors has to be stored if BiCGStab is preconditioned with a multigrid V-cycle.
\section{Implementation}\label{sec:implementation}
As described in \cite{Adams2018}, the LFRic code is designed around a separation-of-concerns philosophy originally introduced in this context in \cite{Ford2013}. It provides well-defined abstractions for isolating high-level science code from computational issues related to low-level optimisation and parallelisation with the aim of achieving performance-portability on different parallel hardware platforms. The model developer (atmospheric scientist or numerical algorithm specialist) writes two kinds of code:
\begin{itemize}
\item \textit{Local kernels}, which describe the operations which are executed in one vertical column of the mesh.
\item High level \textit{algorithms} which orchestrate the kernel calls.
\end{itemize}
The PSyclone code generation system (see \cite{Ford2019}) automatically generates optimised wrapper subroutines for the parallel execution of the kernels over the grid. PSyclone can generate code for execution on distributed memory machines via MPI and shared memory parallelisation with OpenMP, as well as mixed-mode parallelisation; it also supports threaded implementations on GPUs.
Depending on the data dependencies, which are specified by attaching access descriptors to the kernels, appropriate MPI calls (e.g. for halo-exchanges) are automatically inserted into the generated code.

As is common in atmospheric modelling codes and consistent with the tensor-product multigrid algorithm described in the paper, the grid is only partitioned in the horizontal direction, and the elementary kernels operate on contiguous data stored in individual vertical columns. By using a structured memory layout in the vertical direction, any costs of indirect addressing in the horizontal direction from logically unstructured grids can be hidden; this was already observed in \cite{Macdonald2011} and confirmed for tensor-multigrid solvers in \cite{Dedner2016}.

To implement the solvers described in this paper, the user will have to write the iterative solver algorithm and kernels for applying the appropriate finite element matrices or carrying out the block-tridiagonal solves in each column.
\subsection{Solver API}\label{sec:solver-api}
For the complex solvers and preconditioners described in this paper the parameter space is very large: different Krylov solvers for the outer mixed system in Eq. \eqref{eqn:QN_linear_system} and for the Helmholtz pressure equation in Eq. \eqref{eqn:helmholtz} will lead to varying overall runtimes. The performance of the multigrid preconditioner depends on the number of levels, value over-relaxation parameter, the number of pre- and post-smoothing steps and the choice of the coarse level solver.
To explore different configurations and allow the easy switching of components, an object oriented framework for iterative solvers has been implemented in LFRic as described in\cite{Adams2018}. Similar to the DUNE Iterative Solver Template Library (\cite{Blatt2006}) and PETSc (\cite{Balay1997}), this avoids re-implementation of the iterative solver and aids reproducibility. Based on this library of iterative solvers, the user has to provide problem specific linear operators and preconditioner objects.

For this, three abstract data types are defined in the code:
\begin{itemize}
\item A \textit{vector type} which supports linear algebra operations such as \texttt{axpy} updates ($\vec{y}\mapsto \vec{y}+\alpha \vec{x}$) and dot-products ($\beta = \langle \vec{x},\vec{y}\rangle=\sum_j x_j y_j$)
\item A \textit{linear operator type} which acts on vectors $\vec{x}$ and implements the operation $\vec{y}\mapsto A\vec{x}$
\item A \textit{preconditioner type} which implements the operation $\vec{x}\mapsto P\vec{y}$, where $\vec{x}$ approximately solves the equation $A\vec{x}=\vec{y}$
\end{itemize}
This allows the implementation of different Krylov subspace solvers, which are parametrised over the linear operator $A$ and corresponding preconditioner $P$, both of which are derived from their respective abstract base types. So far the following  solvers of the general form \textsf{K($A$,$P$,[$\epsilon$])} (where $\epsilon$ is a tolerance on the target residual reduction) are available in LFRic:
\begin{itemize}
\item Conjugate Gradient \textsf{CG($A$,$P$,$\epsilon$)}
\item Generalised minimal residual \textsf{GMRES($A$,$P$,$\epsilon$)}
\item Stabilised Biconjugate Gradient \textsf{BiCGStab($A$,$P$,$\epsilon$)}
\item Generalised Conjugate Residual \textsf{GCR}($A$,$P$,$\epsilon$)
\item A null-solver (preconditioner only) \textsf{PreOnly($A$,$P$)}
\end{itemize}
All solvers operate on instances of a concrete \textsf{field\_vector} type, which is derived from the abstract base vector type and contains a collection of dof-vectors. More specifically, to implement the linear solver with Schur-complement preconditioner for the linear system described in Section \ref{sec:linear_system}, an operator $A_{\text{mixed}}$ which represents the matrix in Eq. \eqref{eqn:linear_matrix} and acts on a \textsf{field\_vector} for the state $\widetilde{\vec{x}}^{\prime}=(\widetilde{u}^{\prime},\widetilde{\rho}^{\prime},\widetilde{\theta}^{\prime},\widetilde{\Pi}^{\prime})$ was created. The corresponding Schur complement preconditioner $P_{\text{mixed}}$ acts on \textsf{field\_vector}s of the same form and, as discussed in Section \ref{sec:schur_complement}, contains a call to a Krylov-subspace method $\textsf{K}_H(A_H,P_H)$ for solving the Helmholtz problem in Eq. \eqref{eqn:helmholtz}. Here the operator $A_H$ represents the Helmholtz operator in Eq. \eqref{eqn:helmholtz_operator} and $P_H$ is a preconditioner; both $A_H$ and $P_H$ act on a single-component \textsf{field\_vector} objects of the form $\widetilde{\vec{x}}_H^\prime = (\widetilde{\Pi}^{\prime})$. Both the multigrid preconditioner $P_H^{\text{(MG)}}(L,\omega,n_{\text{pre}},n_{\text{post}})$ described in Section \ref{sec:multigrid} and a single-level method $P_H^{(\text{Jac})}(\omega,n_{\text{Jac}})$, which corresponds to $n_{\text{Jac}}$ applications of the block-Jacobi iteration in Eq. \eqref{eqn:jacobi}, were implemented.

Hence the general nested solver can be written as
\begin{equation}
  \textsf{K}_{\text{mixed}}\left(A_{\text{mixed}},P_{\text{mixed}}\left(\textsf{K}_H(A_H,P_H,\epsilon_H)\right),\epsilon\right).\label{eqn:solver_notation}
\end{equation}
\section{Results}\label{sec:results}
To identify the most promising preconditioner, first the performance of different solvers for the pressure correction in Eq. \eqref{eqn:helmholtz} is explored on a relatively small number of compute cores in Section \ref{sec:results_algorithmic}, before presenting massively parallel scaling tests for a smaller subset of solver configurations in Section \ref{sec:results_parallel}. Finally, robustness with respect to the timestep size is quantified in Section \ref{sec:robustness}. All tests were run on the Met Office CrayXC40 supercomputer using the Aries interconnect. Each node is comprised of dual socket, 18-core Broadwell Intel Xeon processors, {\em i.e.} 36 CPU cores per node. The model was compiled with the Intel 17 Fortran compiler (version 17.0.0.098).
\subsection{Algorithmic performance and pressure solver comparison}\label{sec:results_algorithmic}
In all cases a GCR solver with a tolerance of $\epsilon=10^{-6}$ is used to solve the mixed system in Eq. \eqref{eqn:QN_linear_system}; sometimes this will also be referred to as the ``outer solve'' below. Although other methods are available for this option, an investigation of the mixed solver is not the focus of this paper and so only this method, as used in \cite{Melvin2018}, is considered.
\begin{table}
  \begin{center}
    \begin{tabular}{|l|l|l|l|}
      \hline
      pressure solver & $\textsf{K}_H$ & $\epsilon_H$ & $P_H$\\
      \hline\hline
      line relaxation & \textsf{PreOnly} & --- & $P_H^{\text{(Jac)}}(0.8,10)$\\
      \hline
      Krylov ($\epsilon_H$) & \textsf{BiCGStab} & $10^{-2}$ & $P_H^{\text{(Jac)}}(1.0,1)$\\
      \ditto & \ditto & $10^{-4}$ & \ditto \\
      \ditto  & \ditto & $10^{-6}$ &\ditto \\
       \hline
       MG($L$) & \textsf{PreOnly} & --- & $P_H^{\text{(MG)}}(1,0.8,2,2)$\\
      \ditto & \ditto & --- & $P_H^{\text{(MG)}}(2,0.8,2,2)$\\
       \ditto & \ditto & --- & $P_H^{\text{(MG)}}(3,0.8,2,2)$\\
       \ditto & \ditto & --- & $P_H^{\text{(MG)}}(4,0.8,2,2)$\\
       \hline
       Krylov-MG ($\epsilon_H,L$) & \textsf{BiCGStab} & $10^{-2}$ & $P_H^{\text{(MG)}}(4,0.8,2,2)$\\
       \ditto & \ditto & $10^{-4}$ & \ditto\\
       \ditto &\ditto & $10^{-6}$ & \ditto\\
      \hline
    \end{tabular}
  \end{center}
  \caption{Pressure solver configurations used in Section \ref{sec:results_algorithmic}}
  \label{tab:solver_configurations}
\end{table}
To test the algorithmic performance of the solver the model is run on the baroclinic wave test case of \cite{ullrich_etal_12}, which models the development of mid-latitude atmospheric wave dynamics. Apart from the semi-implicit solver, the model setup is the same as described in  \cite{Melvin2018} with the following exceptions:
\begin{enumerate}
\item To improve long timestep stability the continuity equation is handled in an (iterated) implicit manner, instead of explicit as in  \cite{Melvin2018}, that is their (22) becomes \begin{equation} \langle\sigma,\delta_t\rho\rangle = -\Delta t\langle\sigma,\nabla\cdot\overline{\mathbf{F}}^{\alpha}\rangle,\label{eq:si-continuity}\end{equation} with $\overline{\mathbf{F}}^{\alpha}\equiv\alpha\mathbf{u}^{n+1}\rho^{n+1}+\left(1-\alpha\right)\mathbf{u}^{n}\rho^{n}$ sampled pointwise.
\item To improve the accuracy of the advection operator over non-uniform meshes a two-dimensional horizontal polynomial reconstruction of the potential temperature field is used instead of the one-dimensional reconstruction of \cite{Melvin2018}. This reconstruction follows the method of \cite{Thuburn2012}, except here the polynomial is always evaluated at fixed points in space instead of at Gauss points of the swept area as in \cite{Thuburn2012}.
\end{enumerate}

The model is run on a C192 mesh ($6\times192\times192$ cells) with $\approx 50 \rm{km}$ horizontal resolution and 30 levels in the vertical following a quadratic stretching such that the smallest vertical grid spacing is $\approx 200 \rm{m}$. This results in $39.3\cdot 10^6$ total degrees of freedom, with $6.6\cdot 10^6$ pressure unknowns. The timestep is $\Delta t = 1200 \rm{s}$, which results in a horizontal wave Courant number of $\text{CFL}_{\text{h}}=c_s\Delta t/\Delta x\approx 7.9$ with $c_s=340\,\rm{m/s}$. The vertical Courant number is $\text{CFL}_{\text{v}}=c_s\Delta t/\Delta z\approx 1800$ (near the surface).

Different methods are used to solve the pressure correction equation in Eq. \eqref{eqn:helmholtz}:
\begin{enumerate}
  \setlength{\itemsep}{0ex}
\item\textbf{line-relaxation}: 10 iterations of the block-Jacobi solver
\item\textbf{MG($\boldsymbol{L}$)}: Single geometric multigrid V-cycle with varying number of levels $L=1,2,3,4$ and block-Jacobi line smoother on each level.
\item\textbf{Krylov($\boldsymbol{\epsilon_H}$)} and \textbf{Krylov-MG($\boldsymbol{\epsilon_H,L}$)}: BiCGstab iteration with a relative tolerance of $\epsilon_H = 10^{-2}, 10^{-3}, 10^{-6}$ and one of the following preconditioners:
  \begin{enumerate}
    \item One iteration of the block-Jacobi method [Krylov($\epsilon_H$)]
    \item Single geometric multigrid V-cycle with $L=4$ levels [Krylov-MG($\epsilon_H,L$)]
  \end{enumerate}
\end{enumerate}
Following the notation in Eq. \eqref{eqn:solver_notation}, the solver configurations are summarised in Table \ref{tab:solver_configurations}. The Krylov-solver with $\epsilon_H=10^{-6}$ corresponds to the solver setup used in \cite{Melvin2018}. Two pre- and post-smoothing steps with an over-relaxation parameter $\omega=0.8$ are used in the multigrid algorithm; the number of smoothing steps on the coarsest level is $n_{\text{coarse}}=4$.

The test is run for 8 days of simulation time (768 timesteps) on 64 nodes of the CrayXC40 with 6 MPI ranks per node and 6 OpenMP threads per rank.

The accuracy of the linear system is governed by the solution obtained to \eqref{eqn:linear_matrix}, which in all cases uses the same GCR solver and relative tolerance $\epsilon=10^{-6}$. Therefore there is very little difference in the solutions obtained from the different solver setups in Table \ref{tab:solver_configurations}, with the maximum difference in surface pressure after 8 days of simulation being $\approx 0.0001\%$ compared to the Krylov$\left(10^{-6}\right)$ configuration.

\begin{figure}
  \begin{center}
    \includegraphics[width=0.65\linewidth]{\figdir/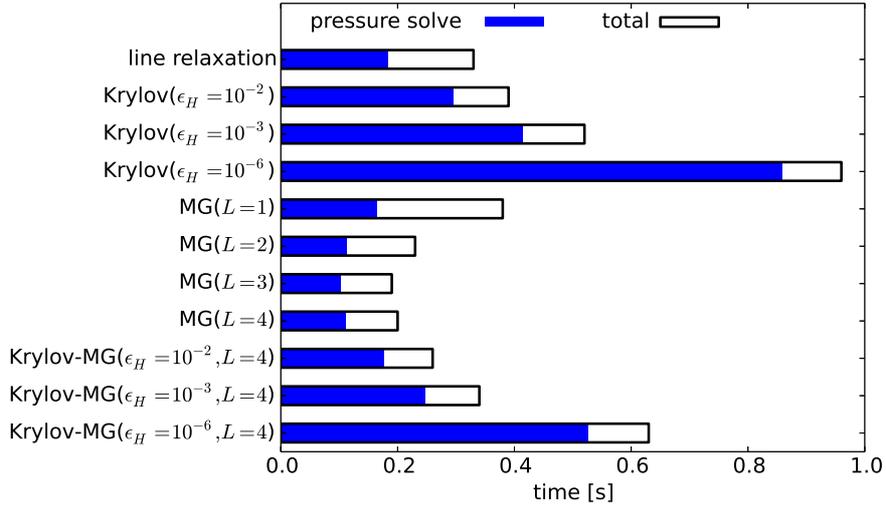}
    \caption{Breakdown of solver times for the baroclinic wave test case on the C192 mesh described in Section \ref{sec:results_algorithmic}. Both the total time per solve and the time spent in the pressure solver (blue) is shown.}\label{fig:solver_timing}
  \end{center}
\end{figure}

Table \ref{tab:solver-timing-summary} lists the average number of outer, mixed solver iterations per semi-implicit solve and the average number of iterations per pressure solve for each of the setups described above and summarised in Table \ref{tab:solver_configurations}. Note that in each timestep the mixed solver is called four times to solve a non-linear problem and Table \ref{tab:solver-timing-summary} shows the average times for a \textit{single} linear solve; those times are visualised in Figure \ref{fig:solver_timing}. For completeness, results for the 1-level multigrid method are also reported. Although in essence this simply corresponds to four applications of the line smoother, since $n_{\text{pre}}+n_{\text{post}}=4$ this guarantees that the same number of fine-level smoother iterations is used for all multigrid results. The multigrid methods (as long as more than one level is used) result in a significant reduction in the time taken for each linear solve and, compared to the Krylov methods, require roughly the same number of outer iterations. In particular, solving the pressure correction equation to a relatively tight tolerance of $\epsilon_H=10^{-6}$ does not reduce the number of outer GCR iterations. This implies that the main error in the approximate Schur complement solve is due to mass lumping, and not due an inexact solution of the pressure equation. Overall, a single multigrid V-cycle with $L=3$ levels gives the best performance. Increasing the number of multigrid levels further does not provide any advantage since the problem is already well-conditioned after two coarsening steps, and adding further coarse levels will make the method slightly more expensive. Although standalone line relaxation provides a cheap pressure solver (without any global sums) this is offset by a significant increase in the number of outer iterations such that the overall cost is not competitive with the multigrid method. In fact (looking at the final column of Table \ref{tab:solver-timing-summary}), 10 iterations of the block-Jacobi solver are slightly more expensive than the four-level multigrid V-cycle with two pre-/post-smoothing steps; this is not too far off the theoretical cost estimate in Eq. \eqref{eqn:MG_cost_estimate}. As the results for the Krylov-MG show, there is no advantage in wrapping the multigrid V-cycle in a Krylov solver. 
\begin{table}
  \begin{center}
  \begin{tabular}{|l|cc|cc|l|}
    \hline
    & \multicolumn{2}{|c|}{mixed solve}
    & \multicolumn{2}{|c|}{pressure solve} & solver\\
    pressure solver & \# iter & \tmixed & \# iter & \tpress & setup \\
    \hline\hline
    line relaxation & 24.54 & 0.33 & 10 & 0.0075 & 0.018\\
    \hline
    Krylov($10^{-2}$) & 15.10 & 0.39 & 15.12 & 0.0196 & 0.020\\
    Krylov($10^{-3}$) & 15.03 & 0.52 & 22.59 & 0.0276 & 0.019\\
    Krylov($10^{-6}$) & 14.03 & 0.96 & 52.50 & 0.0571 & 0.018\\
    \hline
    MG($1$) & 35.29 & 0.38 & --- & 0.0047 & 0.018\\
    MG($2$) & 19.51 & 0.23 & --- & 0.0058 & 0.027\\
    MG($3$) & 15.24 & 0.19 & --- & 0.0067 & 0.026\\
    MG($4$) & 15.12 & 0.20 & --- & 0.0073 & 0.026\\
    \hline
    Kr-MG($10^{-2},4$) & 15.04 & 0.26 & 1.58 & 0.0117 & 0.028\\
    Kr-MG($10^{-3},4$) & 15.03 & 0.34 & 2.21 & 0.0165 & 0.026\\
    Kr-MG($10^{-6},4$) & 15.03 & 0.63 & 4.37 & 0.0350 & 0.026\\
    \hline
  \end{tabular}
  \end{center}
  \centering{}\protect\caption{\label{tab:solver-timing-summary}
Number of iterations and time per linear solve for both the outer mixed solve (\tmixed) and for the inner pressure solve (\tpress). Setup time per mixed solve for the linear operators is given in the final column. All numbers are averaged over the total run of the baroclinic wave test case on the C192 mesh described in Section \ref{sec:results_algorithmic}; times are given in seconds. To improve formatting, ``Krylov-MG'' has been abbreviated to ``Kr-MG'' in the final three rows.}
\end{table}

Although the use of the multigrid preconditioner significantly improves speed of the linear solver it does come with additional setup costs. Principally these come from two areas that are not covered in the solver timings in Figure \ref{fig:solver_timing}, the first is setup costs of the multigrid hierarchy, such as reading in the multiple meshes, computing the intergrid mappings and computation of certain temporally invariant operators, such as the mass matrices on every grid level. The additional cost of this area is marginal and does not show itself in the overall model runtimes for the tests in Figure \ref{fig:solver_timing}. The second area is the computation of operators in \eqref{eqn:Helmholtz_operator} that depend upon the reference state (those denoted with a $*$ in \eqref{eqn:Helmholtz_operator}) and the restriction of the reference state $\left(\Pi^*,\, \rho^*,\,\theta^*\right)$ between different levels of the multigrid hierarchy. The average cost per mixed solve of computing these operators for all solver configurations is given in the final column of Table \ref{tab:solver-timing-summary}, if any multigrid levels are used then the cost of computing these operators increases by $\approx 50\%$, but adding more levels after the first does not alter the cost significantly. This increase is not a significant fraction of the overall model runtime and is more than outweighed by the savings from the solver, additionally all these computations are local to the processing element and so would be expected to scale well.

Results (not shown) using the alternative iterative methods listed in Section \ref{sec:solver-api} for the pressure solver with $\epsilon_H=10^{-2}$ show similar performance to the BiCGstab solver presented in Table \ref{tab:solver-timing-summary}. The only exceptions are the Jacobi \& precondition-only methods which take approximately twice as long to run as using BiCGstab for the pressure solver. Since the linear problem is not symmetric, the Conjugate Gradient method does not converge.
\subsection{Massively parallel scalability}\label{sec:results_parallel}
To examine parallel scalability, two representative pressure solvers are chosen, namely BiCGStab with a prescribed tolerance of $\epsilon_H = 10^{-2}$ (denoted ``Kr2'' hereafter) and a standalone multigrid V-cycle with 3-levels (denoted ``MG''). For comparison, BiCGstab with a prescribed tolerance of $\epsilon_H = 10^{-6}$ (denoted ``Kr6'') is also included in the following study. The scaling tests are run on a C1152 cubed sphere mesh ($6\times 1152\times 1152$ cells) with $\approx 9$ km horizontal resolution and 30 vertical levels with the same stretched grid and vertical $\rm{CFL}_v$ as in Section \ref{sec:results_algorithmic}. The total number of degrees of freedom is $1.4\cdot 10^9$ with $2.4\cdot10^8$ pressure unknowns. The timestep is set to $205$ seconds, such that the horizontal wave Courant number is again $\rm{CFL}_h=c_s\Delta t/\Delta x\approx 8.0$,
as in Section \ref{sec:results_algorithmic}. In a strong scaling experiment the model is run for 400 timesteps\footnote{Due to time constraints, the model is only run for only 250 time steps with the Kr6 solver on 384 nodes.} on up to 3456 nodes of the Cray XC40, keeping the global problem size fixed. Each node is configured to run with 6 MPI ranks and 6 OpenMP threads per rank; for the largest node count this corresponds to $3456\times 36 = 124416$ threads. In this case the local state vector has around $10000$ degrees of freedom and there are only around $2000$ pressure unknowns per core. In~\cite{Fischer2016}, an analysis of solver algorithms is combined with a performance model for the computation and communication costs. According to this analysis, strong scaling is expected to break down when the local number of unknowns is smaller than $\mathcal{O}(10^4 - 10^5)$. It is therefore expected that the largest node count considered in the present paper is well in the strong scaling limit. Table \ref{tab:problemsizes} summarises the resulting local problem sizes for all node counts.

The average time spent in the mixed- and pressure- solver are shown in Table \ref{tab:scale}. Note that those times are now reported for an entire timestep, i.e. aggregated over four linear solves. This will allow a quantitative comparison to the overall communication cost per timestep reported below.
\begin{table}
  \begin{center}
    \begin{tabular}{|r|r|r|rr|}
      \hline
      \# nodes & \# threads & \multicolumn{1}{c|}{\# local} & \multicolumn{2}{c|}{\# local unknowns}\\
      & & \multicolumn{1}{c|}{columns} & mixed & pressure\\
      \hline\hline
      $384$ & $13824$ & $24\times 24$ & $103968$ & $17280$\\
      $864$ & $31104$ & $16\times 16$ & $46528$ & $7680$ \\
      $1536$ & $55296$ & $12\times 12$ & $26352$ & $4320$ \\
      $3456$ & $124416$ & $8\times 8$ & $11872$ & $1920$\\
      \hline
    \end{tabular}
    \caption{Node configurations and local problem sizes for parallel strong scaling runs in Section \ref{sec:results_parallel}.}\label{tab:problemsizes}
\end{center}
\end{table}

\begin{table}
  \begin{center}
\begin{tabular}{|r|rr|rr|rr|}
  \hline
nodes  & \multicolumn{2}{|c|}{MG} & \multicolumn{2}{|c|}{Kr2} & \multicolumn{2}{|c|}{Kr6} \\
         & \Tmixed  & \Tpress & \Tmixed  & \Tpress & \Tmixed  & \Tpress \\ \hline\hline
$384$  & $4.70$ & $2.34$    & $15.1$ & $10.3$    & $35.0$ & $26.3$\\
$864$  & $3.14$ & $1.30$    & $8.82$  & $5.52$   & $14.0$  & $10.9$\\
$1536$ & $1.63$  & $0.563$    & $5.96$  & $3.69$   & $16.7$  & $10.5$\\
$3456$ & $1.15$  & $0.285$    & $5.71$  & $3.34$   & $11.1$  & $7.24$\\\hline
\end{tabular}
\end{center}
\centering{}\protect\caption{\label{tab:scale}Strong scaling of the time spent in the linear solver on different numbers of nodes for the C1152 mesh test case in Section \ref{sec:results_parallel}. All results are measured in seconds and are averaged over 400 timesteps. \Tmixed~ denotes the average total time spent in the mixed, outer solve for each timestep; \Tpress~ is the average time spent in the inner pressure solve.}
\end{table}

It should be noted that the Aries network deployed on the Cray XC40 uses an adaptive routing algorithm for the messages~(\cite{Roweth2019}). This ensures the best utilisation of all the network links across the machine, but may not be optimal for a given task. Moreover, the path taken by messages is not the same each time resulting in variation in the time taken. Ideally, a statistical measure, such as the minimum or mean would be taken over many measurements to discount such variation. However, jobs running on such large numbers of nodes are computationally expensive and the large number of potential paths through the network would require a large (and expensive) statistical sample, which was not feasible for this study.  Consequently, only a single result for each setup is reported.

\begin{figure}
  \begin{center}
    \includegraphics[width=0.6\linewidth]{\figdir/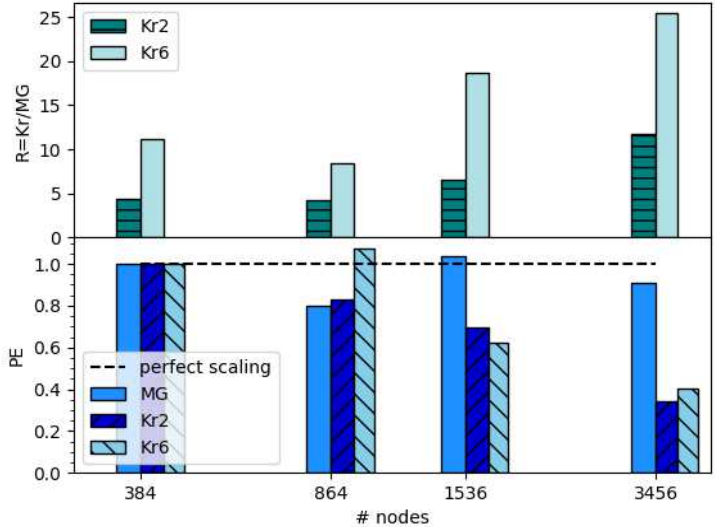}
    \caption{\label{fig:scale_pressure}Strong scaling of the time spent in the inner pressure solver for the C1152 mesh test case in Section \ref{sec:results_parallel}. The lower panel shows the parallel efficiency $\text{PE}$ compared to 384 nodes as defined in Eq. \eqref{eqn:PE}. The horizontal dashed line corresponds to perfect scaling where $\text{PE}=1$. The relative cost of the Krylov solvers (Kr2 and Kr6) compared to  the multigrid V-cycle (MG) is shown at the top.}
  \end{center}
\end{figure}
Figure~\ref{fig:scale_pressure} quantifies the strong scaling of the time spent in the inner pressure solver. Rather than plotting the absolute times (which can be read off from Table \ref{tab:scale}), the parallel efficiency and the relative cost of the Krylov-subspace solvers (Kr2, Kr6) compared to multigrid (MG) are shown. Let $\Tpress(N)$ be the time spent in the pressure solve on $N$ nodes. The parallel efficiency $\text{PE}$ relative to $N_{\text{ref}}=384$ nodes is defined as
\begin{equation}
  \text{PE} = \frac{\Tpress(N_{\text{ref}})/\Tpress(N)}{N/N_{\text{ref}}}.
  \label{eqn:PE}
\end{equation}
As the medium blue, unhatched bars show, the multigrid solver scales extremely well to 3456 nodes.
Both Krylov solvers (with the dark blue \hatchingNW\ hatching for Kr2 and light \hatchingNE~ hatching for Kr6) show significantly worse scaling; that the trend is not smooth can be attributed to network variability. The upper panel of Figure \ref{fig:scale_pressure} shows the relative performance of the multigrid solver and the two Krylov methods for increasing node counts. More specifically, the ratio $\text{R}=\text{Kr}/\text{MG}$ is obtained by dividing the time spent in one of the Krylov solvers by the time for the multigrid solver. The MG solver is much faster than either Kr2 and Kr6, especially for large node counts where the superior scalability of the multigrid algorithm pays off: for 3456 nodes, the MG solver is more than $11\times$ faster than Kr2 and $25\times$ faster than Kr6. As will be discussed below, this can be at least partially explained by the fact that multigrid does not rely on costly global reductions which are required in each iteration of the Krylov-subspace solvers. However, even for 384 nodes the relative advantage of MG is more than $4\times$ compared to the Kr2 solver and $11\times$ compared to Kr6.
\begin{figure}
  \begin{center}
    \includegraphics[width=0.6\linewidth]{\figdir/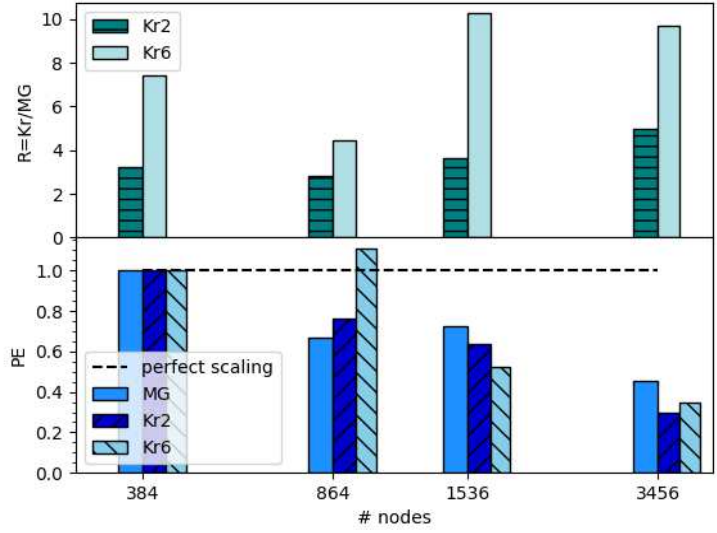}
    \caption{\label{fig:scale_si}Strong scaling of the time spent in the outer, mixed solver for the C1152 mesh test case in Section \ref{sec:results_parallel}. The lower panel shows the parallel efficiency $\text{PE}$ compared to 384 nodes as defined in Eq. \eqref{eqn:PE}. The horizontal hashed line shows perfect scaling where $\text{PE}=1$. The relative cost of the Krylov pressure solvers (Kr2 and Kr6) compared to  the multigrid V-cycle (MG) is shown at the top.}
  \end{center}
\end{figure}

Figure \ref{fig:scale_si} shows the same quantities as in Figure \ref{fig:scale_pressure}, but now for the outer, mixed solve. Again the parallel efficiency is given in the lower panel and the upper panel quantifies the relative advantage of the MG pressure solver, compared to Kr2 and Kr6. 
Compared to Figure \ref{fig:scale_pressure}, the strong parallel efficiency of the outer solve drops to less than $40\%$ for all pressure solvers. This can be explained by the additional parallel communications in the outer solver, which have a particularly strong impact for MG. As can be seen from Table \ref{tab:scale}
 this is to be expected since the multigrid solver accounts for a relatively smaller fraction of the outer solve time. However, as can be read off from the top panel of Figure~\ref{fig:scale_si}, the MG pressure solver still improves performance relative to Kr2 and Kr6: on the largest node count multigrid leads to a $4\times$ reduction in runtime compared to Kr2 ($9\times$ for Kr6) and even on 384 nodes the relative advantage of MG is $3\times$ for Kr2 ($7\times$ for Kr6).
\begin{figure}
  \begin{center}
    \includegraphics[width=0.6\linewidth]{\figdir/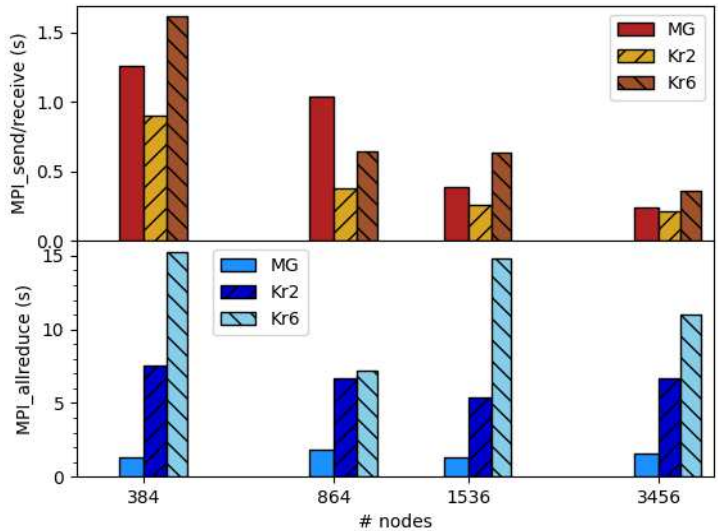}
    \caption{\label{fig:scale_comms}Strong scaling of the average communication costs per timestep for the C1152 test case in Section \ref{sec:results_parallel}. The upper panel shows the cost of MPI send/receive operations during halo-exchange; lower panel shows the time spent in global sums. All times are measured with the CrayPAT profiling tool.}
  \end{center}
\end{figure}

To understand the differences in parallel scalability for the different pressure solvers, Figure~\ref{fig:scale_comms} shows the communication costs for increasing numbers of nodes. This includes both local communications in halo exchanges (top panel) and all-to-all communications in global reductions (bottom panel). The numbers were collected with the CrayPAT profiler in sampling mode. For technical reasons it was not possible to limit the measurements to the solver routine. Instead, the measured data is aggregated across all the calls to the relevant MPI library functions for the whole model run. However, from the profiling data, the time spent in the semi-implicit solver varies from $58-72\%$ of the time per timestep for MG, $79-87\%$ for Kr2 and $89-92\%$ for Kr6. Moreover almost all the calls to the global sum take place in the solver routines. Thus it is reasonable to conclude that the communication costs are dominated by the solver.

The lower panel in Figure~\ref{fig:scale_comms} shows $\tau$, the time spent in the global sums per timestep. On the largest node count, the approximate times are $\tau\approx 2\rm{s}$ for MG (blue, unhatched), $\tau\approx 7\rm{s}$ for Kr2 (dark blue with \hatchingNE~ hatching) and $\tau\approx 12\rm{s}$ for Kr6 (light blue with \hatchingNW~ hatching). Evidently Kr2 and Kr6 spend much more time in global communications. This is readily explained by the fact that the number of global sums in the pressure solve is proportional to the number of Krylov solver iterations. In contrast, the MG V-cycle does not require any global sums and the superior scaling of the MG pressure solver itself (Figure \ref{fig:scale_pressure}) can be largely accounted for by this absence of the global communication. Moreover, from Figure~\ref{fig:scale_comms} it can be seen that especially for Kr6 the variation in the cost of the global sum is large. To quantify this, define the variation $\nu_t$ as
\begin{equation}
  \nu_t = \frac{t_{\rm max} - t_{\rm min}}{t_{\rm max} + t_{\rm min}},
\end{equation}
where $t_{\rm min}$ and $t_{\rm max}$ are the smallest and largest measured global communication times across all four considered node counts.
For the different methods the numerical values are $\nu_t=0.18$ for MG, $\nu_t=0.17$ for Kr2 and $\nu_t=0.36$ for Kr6. Apart from the longer time to solution, the large variation in global communication costs due to network variability is another disadvantage of the Kr6 method. Again, this can be explained by the significant fraction of time spent in global sums during the pressure solve. For MG, global sums are only required in the outer solve and consequently the run-time variation is much smaller.

Although the trend in the data is not entirely clear due to the network variability, the Kr2 solver (yellow with \hatchingNE~ hatching) spends the least amount of time in nearest-neighbour communications. Since Kr6 (brown with \hatchingNW~ hatching) requires more iterations to converge, it will also require more halo exchanges. The nearest-neighbour communication cost of the MG V-cycle (red) lies somewhere between those two extremes, which can be attributed to additional halo exchanges on the coarser multigrid levels. Note, however, that limiting the number of levels to $L=3$ in the shallow multigrid approach avoids the inclusion of very coarse levels with a poor computation-to-communication ratio. The cost of halo exchange decreases with the number of nodes. This is again plausible since (as long as the message size is not too small) local communication is bandwidth bound and so scales with the amount of data that is sent. The third column of Table \ref{tab:problemsizes} shows that the size of the halo reduces by a factor of three as the number of nodes increases from 384 to 3456.

The Multigrid solver scales better than might be expected from~\cite{Fischer2016} which predicts that for the largest node count the local problems are so small that strong scaling should break down. This can be at least partly attributed to the fact that the analysis in~\cite{Fischer2016} is for a three-dimensional decomposition, whereas in the present work the decomposition is only two-dimensional. Surface-to-volume scaling implies that for small local volumes, there is less communication in 2-d than 3-d.

As discussed in section \ref{sec:results_algorithmic} there are extra costs for the set up of the multigrid solver.
The dominant costs arise from the computation of operators, in (\ref{eqn:Helmholtz_operator}), that depend on the reference state and the restriction of the reference state between levels in the multigrid hierarchy. The cost of this computation is small compared to the overall run-times, in some cases less than $1\%$, which is then not automatically captured in the profile. However, from the incomplete data captured it is apparent that firstly, the computation of the operators scales very well with processor number (as it is proportional to the problem size). Secondly, the excess cost for multigrid is small, roughly a $50\%$ increase compared to Krylov and moreover, the excess is itself small compared to the cost of the multigrid pressure solve, about $6\%-8\%$.

\subsection{Robustness with respect to timestep size}\label{sec:robustness}
Finally, the impact of variations in the horizontal acoustic Courant number $\text{CFL}_{\text{h}}$ on the performance of the model is studied. As can be inferred from the analysis in Section \ref{eqn:helmholtz_structure}, the condition number of the Helmholtz operator $H$ in Eq. \eqref{eqn:helmholtz_operator} depends strongly on $\text{CFL}_{\text{h}}$. In other words, for a fixed grid spacing the condition number will increase with the timestep size, which can potentially make the pressure solver more costly. To quantify the impact of this, the model was run on 384 nodes with timestep sizes of varying length. Apart from this, the setup is the same as in Section \ref{sec:results_parallel}. Table \ref{tab:CFL} shows the number of outer, mixed and pressure solver iterations for $\text{CFL}_{\text{h}}$ between 4 and 8. Results for $\text{CFL}_{\text{h}}=4, 6$ are averaged over the first 100 timesteps. The final column of Table \ref{tab:CFL} shows the average time spent in the linear solver in each timestep. Unfortunately it was not possible to chose larger timestep sizes due to instabilities due to the CFL limit imposed by the explicit advection scheme, (for the simulations in Section \ref{sec:results_algorithmic} with the horizontal wave $\text{CFL}_{\text{h}}\approx 8$, the advective Courant number grows from $\text{CFL}_{\text{adv,h}}\approx 0.6$ early in the simulation to around $\text{CFL}_{\text{adv,h}}\approx 1.3$ and the end of the simulation; the vertical advective Courant number is $\text{CFL}_{\text{adv,v}}\approx 1.8$ by the end of the simulation). For the Kr2 and Kr6 solvers the number of inner, pressure iterations is also given. It can be observed that the number of outer iterations is largely insensitive to increases in the timestep size, in particular using one multigrid V-cycle for the pressure solve gives good results for the range of $\text{CFL}_{\text{h}}$ considered here. Any increase in the wall-clock time for the configurations which use Krylov-subspace pressure solvers can be clearly attributed to the growing number of inner iterations. Relative to the MG setup, which only requires exactly one multigrid V-cycle in the pressure solve, the cost of the Kr2 and Kr6 solvers increases for larger values of $\text{CFL}_{\text{h}}$. This is readily explained by the fact that the number of pressure solver iterations grows with the condition number. Empirically it can be observed that this number of iterations approximately doubles when $\text{CFL}_{\text{h}}$ is increased from 4 to 8. While values of $\text{CFL}_{\text{h}}\approx10$ are typical atmospheric simulations, even for smaller timestep sizes the multigrid pressure solver shows a clear advantage compared to Krylov-subspace methods.
\begin{table}
  \begin{center}
\begin{tabular}{|c|c|rr|r|}
\hline
\multirow{2}{*}{$\text{CFL}_{\text{h}}$} & \multirow{2}{*}{Solver} & \multicolumn{2}{c|}{\# iter} & \multirow{2}{*}{$\Tmixed\;$} \\
& & mixed & pressure & \\
\hline\hline
$4$& MG  & $12.9$ & $-$    & $3.96$ \\
$4$ & Kr2 & $13.2$ & $9.4$  & $9.13$ \\
$4$ & Kr6 & $12.0$ & $26.2$ & $16.29$ \\\hline
$6$ & MG  & $13.6$ & $-$    & $4.98$ \\
$6$ & Kr2 & $13.3$ & $13.2$ & $10.31$ \\
$6$ & Kr6 & $12.3$ & $40.4$ & $22.16$ \\\hline
$8$ & MG  & $14.0$ & $-$    & $4.70$ \\
$8$ & Kr2 & $13.3$ & $17.3$ & $15.15$ \\
$8$ & Kr6 & $12.6$ & $54.2$ & $34.96$ \\\hline
\end{tabular}
\end{center}
  \centering{}\protect\caption{\label{tab:CFL}Average number of solver iterations for the outer mixed and inner pressure solves for increasing horizontal acoustic Courant number. Results are reported for the C1152 test case in Section \ref{sec:results_parallel}. The final column lists the average time for the spent in the mixed, outer solve for each timestep.}
\end{table}

\section{Conclusion}\label{sec:conclusion}
This paper describes the construction of a Schur-complement preconditioner with a multigrid pressure solver for the mixed-finite element LFRic numerical forecast model. Due to the presence of a velocity mass matrix-matrix an additional outer solver iteration is necessary, which makes the solver significantly more complex than in simpler finite-difference models on structured latitude-longitude grids. By exploiting the structure of the Helmholtz-operator on the highly anisotropic global grid, it is possible to build a highly-efficient bespoke multigrid method using the tensor-product approach in \cite{Borm2001}. Using only a relatively small number of multigrid levels further improves parallel scalability.

The numerical results presented here confirm the conclusions from earlier studies in idealised contexts, as described e.g. in \cite{Mitchell2016}: compared to Krylov-subspace methods, solving the pressure correction equation with a single multigrid V-cycle leads to significantly better overall performance. Running a baroclinic test case with more than 1 billion ($10^9$) unknowns on 124416 threads, multigrid reduces the time spent in the outer, mixed solve by a factor of around $4\times$. Since it requires significantly fewer global reductions, multigrid also scales better in parallel. In contrast to Krylov-subspace based pressure solvers, multigrid is robust with respect to changes in the timestep size.

There are several avenues for future work. While multigrid has been demonstrated to be consistently better than any other considered solver, there are likely further small, problem specific optimisations which can be achieved for example by tuning parameters or including additional terms in the approximate Helmholtz operator by using a modified mass-lumping strategy. While the focus in this paper was on the lowest-order discretisation, \cite{Mitchell2016} have demonstrated that the approach can be easily extended to higher orders by including an additional $p$-refinement step in the multigrid hierarchy. LFRic already provides the necessary code infrastructure. Following the promising results shown here, extending the approach from global simulations to local area models will require careful treatment of boundary conditions but is likely to lead to similar performance gains. Finally, an obvious downside of the approximate Schur-complement approach pursued here is the necessity of an additional outer solve of the mixed problem. As has been recently shown in \cite{Gibson2018}, this can be avoided by using a hybridised mixed finite element discretisation. In fact, for a gravity-wave testcase a solver based on this hybridised approach has already been implemented in LFRic. Although not yet available at the moment, an efficient preconditioner for this solver is actively developed.
\section*{Acknowledgements}
The authors thank the GungHo network (NERC grant NE/K006762/1) for financial support during collaboration visits. Part of this work was carried out during a secondment of EM funded by the Bath Institute for Mathematical Innovation.

\end{document}